\documentclass[a4paper,fleqn,usenatbib]{mnras}

\usepackage[T1]{fontenc}
\usepackage{ae,aecompl}

\usepackage{graphicx}	% Including figure files
\usepackage{amsmath}	% Advanced maths commands
\usepackage{amssymb}	% Extra maths symbols
\usepackage{multirow}
\usepackage{array}
\usepackage{threeparttable}

\title[Spectral classification]{Spectral classification and composites of galaxies in LAMOST DR4}

\author[Li-Li Wang et al.]{
	   Li-Li Wang,$^{1,2,3}$
       A-Li Luo,$^{1}$\thanks{E-mail: lal@nao.cas.cn}
       Shi-Yin Shen,$^{4}$
       Wen Hou,$^{1}$
       Xiao Kong,$^{1}$
       Yi-Han Song,$^{1}$
  \newauthor
       Jian-Nan Zhang,$^{1}$
       Wu Hong,$^{1}$
       Zi-Huang Cao,$^{1}$
       Yong-Hui Hou,$^{5}$
       Yue-Fei Wang,$^{5}$
   \newauthor
       Yong Zhang,$^{5}$
       and Yong-Heng Zhao $^{1}$
\\
$^{1}$Key Laboratory of Optical Astronomy, National Astronomical Observatories, Chinese Academy of Sciences, Beijing 100012, China\\
$^{2}$School of Information Management, Dezhou University, Dezhou 253023, China\\
$^{3}$University of Chinese Academy of Sciences, Beijing 100049, China\\
$^{4}$Key Laboratory for Research in Galaxies and Cosmology, Shanghai Astronomical Observatory, Chinese Academy of Sciences, \\
Shanghai 200030, China\\
$^{5}$Nanjing Institute of Astronomical Optics \& Technology, National Astronomical Observatories, Chinese Academy of Sciences,\\ Nanjing 210042, China\\
}

\date{Accepted XXX. Received YYY; in original form ZZZ}

\pubyear{2015}

\begin{document}
\label{firstpage}
\pagerange{\pageref{firstpage}--\pageref{lastpage}}
\maketitle

\begin{abstract}
We study the classification and composite spectra of galaxy  in the fourth data release (DR4) of the Large Sky Area Multi-Object Fiber Spectroscopic Telescope (LAMOST). We select 40,182 spectra of galaxies from LAMOST DR4, which have photometric information but no spectroscopic observations in the Sloan Digital Sky Survey(SDSS). These newly observed spectra are re-calibrated and classified into six classes, i.e. passive, H$\alpha$-weak, star-forming, composite, LINER and Seyfert using the line intensity (H$\beta$, [OIII]$\lambda$5007, H$\alpha$ and [NII]$\lambda$6585). We also study the correlation between spectral classes and morphological types through three parameters: concentration index, ($ u $ - $ r $) color, and D4000$_n$ index. We calculate composite spectra of high signal-to-noise ratio(S/N) for six spectral classes, and using these composites  we pick out some features that can differentiate the classes effectively, including H$\beta$, Fe5015, H$\gamma_A$, HK, and Mg$_2$ band etc. In addition, we compare our composite spectra with the SDSS ones and analyse their difference. A galaxy catalogue of 40,182 newly observed spectra (36,601 targets) and the composite spectra of the six classes are available online.

\end{abstract}


%%%%%%%%%%%%%%%%%%%%%%%%%%%%%%%%%%%%%%%%%%%%%%%%%%

%%%%%%%%%%%%%%%%% BODY OF PAPER %%%%%%%%%%%%%%%%%%

\section{Introduction}
There are various and complicated phenomena in galaxy formation and evolution. One of the major goals of extragalactic astronomy is to comprehend the nature of the diverse galaxies. The first step towards the goal is to classify them using some criteria and compare their properties among the classes. There are some frequently used criteria for classification: morphology, color and spectral features. The morphological classes include elliptical galaxies, lenticular galaxies, spiral galaxies, barred spiral galaxies and irregular galaxies (\citealt{Hubble1936}, \citealt{Lintott et al2008}, \citealt{Shimasaku et al2001}). The classes by color are blue, green-valley and red (\citealt{Morgan et al1957}, \citealt{Strateva et al2001}, \citealt{Martin et al2007}). Based on spectral features of emission lines, the galaxies are mainly divided into star-forming galaxies and AGNs(\citealt{Baldwin et al1981}, \citealt{Kewley et al2001}, \citealt{Kauffmann et al2003a}, \citealt{Brinchmann et al2004}, \citealt{Kewley et al2006}, \citealt{Stasinska et al2006}, \citealt{Cid Fernandes et al2010}). Some researchers also use more than one criterion, for example, in the work of \cite{Lee et al2008}, galaxies were classified into 16 classes by morphology, color and  spectral features, and \cite{Dobos et al2012} presented a refined classification using both color and spectral features for galaxies. For the Large Sky Area Multi-Object Fiber Spectroscopic Telescope (LAMOST) survey, we use spectral features in galaxy classification in this paper .

Classifications driven by spectral features are focused on emission-line galaxies. The classical classification scheme pioneered by \cite{Baldwin et al1981} (dubbed the BPT diagram) has been widely used over the last three decades. Based on the $\textsc{BPT}$ diagnostic diagram, which uses [OIII]$\lambda$5007/H$\beta$ and [NII]$\lambda$6585/H$\alpha$ line ratios, the emission-line galaxies are classified into star-forming galaxies, composite galaxies, LINERs and Seyferts. There are several empirical segregation curves on $\textsc{BPT}$ diagrams for classification, such as \cite{Kewley et al2001}( hereafter K01), \cite{Kauffmann et al2003a}( hereafter K03) and \cite{Kewley et al2006}( hereafter K06). The curves defined by K03 and K01 represent the border lines of star-forming galaxies and AGNs, and the K06 criteria is used for separating Seyferts and LINERs. There are some other alternative diagnostic diagrams for classification. For example, in \cite{Stasinska et al2006}, the DEW diagram was proposed to distinguish star-forming galaxies and AGNs, using the D4000$_n$ index vs. max(EW[OII], EW[NeIII]). \cite{Cid Fernandes et al2010} introduced a diagram (named WHAN diagram) based on equivalent width of H$\alpha$ vs. the ratio of [NII] and H$\alpha$, which is able to cope with the large population of weak line galaxies in SDSS galaxies.

Galaxies with no or weak signal of emissions are referred to as passive galaxies. In \cite{Lee et al2008}, passive galaxies were selected as galaxies with no or insufficient signal of H$\alpha$ emission. In \cite{Dobos et al2012}, passive galaxies were divided into two further classes: completely passive with no detectable emission lines and passive with weak H$\alpha$ emission. In this paper, we make a classification scheme for LAMOST spectra referring to \cite{Dobos et al2012}.

Composite spectra have been widely applied in researches of extragalactic objects (\citealt{Vanden Berk et al2001}, \citealt{Eisenstein et al2003}, \citealt{Dobos et al2012}). These high signal-to-noise ratio(S/N) composites reveal variations from general continuum and weak emission features that are rarely detectable in individual spectra. There are two methods for stacking spectra: mean spectrum and median spectrum which respectively obtain optimal measurements of the continuum and the emission lines \citep{Vanden Berk et al2001}. Specifically, the mean method further includes geometric mean and algebraic mean. The geometric mean is suitable for averaging the continuum of power-law spectra such as quasars and the algebraic mean is appropriate to the continuum of galaxy spectra, of which the continuum is basically a (linear) superposition of the spectra of various stellar populations. However the mean composite can not guarantee to preserve the real emission line ratios in case that the intensities of emission lines vary significantly in spectral bins \citep{Stasinska et al2015} or there be noises in some emission lines. The median spectrum is robust against noise and preserves the relative fluxes of the emission features, but it might yield non-physical continua because it treats the spectral bins independently. In addition to the methods above, \cite{Yip et al2004}, \cite{Dobos et al2012} compute the composites using principal component analysis, simultaneously dealing with noisy and gappy data where certain spectral bins are masked out due to bad observations or other reasons. The idea of the gap-correction process is to reconstruct the missing regions in the spectrum using its principal components.

With the LAMOST spectroscopic survey going on, a large data set is provided to study galaxies in our nearby universe(\citealt{Luo et al2015}). Several well-known surveys have been carried out such as the Sloan Digital Sky Survey (SDSS)(\citealt{York et al2000}) which observed the largest number of extragalactic targets. In LAMOST DR4, there are tens of thousands of galaxy spectra which have photometric information but no spectroscopic observations in SDSS Data Release Thirteenth, DR13 \citep{Albareti et al2016}. These newly observed spectra in LAMOST DR4 are our study objects in this paper. The primary goals of this paper are to (1) classify these newly observed spectra by using spectral line features(H$\beta$, [OIII]$\lambda$5007, H$\alpha$ and [NII]$\lambda$6585), presenting a catalogue with classification information and more accurate flux measurements of the nebular emission lines, and (2) compute the composite spectra of different classes to extract typical spectral features, and compare our composites with similar composite spectra from the SDSS DR7 (\citealt{Dobos et al2012}).

The outline of this paper is as follows. Section 2 shows the data set we used. Section 3 re-calibrates the fluxes of spectra of galaxies and presents the line measurement of the spectra. In section 4 we describe our classification scheme of galaxies for LAMOST and study the correlation between spectral classes and morphological types. In section 5 we compute the composite spectra of different classes to analyze the global spectral properties, and compare our composites with the SDSS ones. The summary is given in section 6. We assume the cosmological parameters with $ H_0 $=100 km $ s^{-1} $, $ \Omega_M $=0.3, $ \Omega_{\Lambda} $=0.7.

\section{Data}

\subsection{Galaxies in LAMOST DR4}

LAMOST is dedicated to a spectroscopic survey that covers celestial objects over the entire available northern sky. The telescope is characterized by both a large field of view and large aperture, with an effective aperture of 3.6--4.9m and 4,000 fibers mounted on its focal plane. Its spectral wavelength ranges from 3800\AA \ to 9000\AA \, and spectral resolution is about R$\sim$1800 (\citealt{Cui et al2012}, \citealt{Zhao et al2012}). The fourth data release (DR4) of LAMOST includes the pilot survey(2011 October to 2012 June) and the regular survey(2012 September to 2016 June), and will be public released in June 2018. The LAMOST spectroscopic classification system classifies the spectra into STAR, GALAXY, QSO and UNKNOWN. Most broad line AGNs are classified as 'QSO', and some of them are in 'GALAXY'. In this paper, to avoid contaminating our sample with broad line AGNs, we remove the spectra that contain any strong broad line with FWHM>1000km $ \rm s^{-1} $ \citep{Vanden Berk et al2006}. Therefore, a small number(1\%) of galaxy spectra are excluded from our analysis.

\begin{figure}
	\includegraphics[width=9cm]{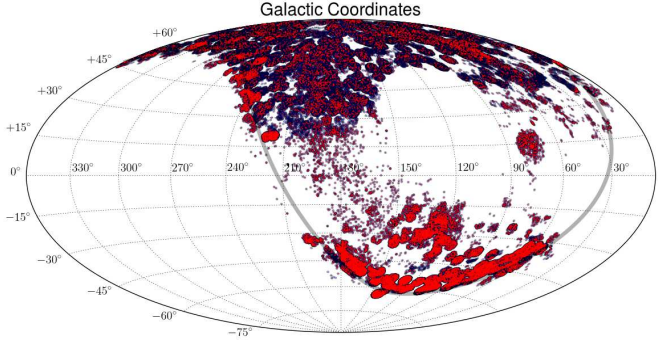}
    \caption{The footprints of all galaxies in $\textsc{LAMOST DR4}$ in the galactic coordinates. The bold grey line indicates Dec=$0^{\circ}$ in declination, and most of the LAMOST galaxies locate in the northern sky. The blue background is the total observations of galaxies in DR4, while the red points represent LAMOST galaxy spectra without spectroscopic observations in SDSS DR13.}
    \label{fig:figure1}
\end{figure}

The footprints of all galaxies in LAMOST DR4 in galactic coordinates are shown in Figure \ref{fig:figure1}. There are two main regions of the extragalactic survey \citep{Luo et al2015}. One is in the Northern Galactic Cap region($b\geq$ $30^{\circ}$), and about 77,154 spectra are observed in this region. The other region is the Southern Galactic Cap region($b\leq$ $-30^{\circ}$), and there are about 32,453 spectra. In the remaining region ($|b|<30^{\circ}$), a small number of 8,899 spectra are obtained. The statistical result is shown in Table \ref{tab:count_spec}.

\begin{table}
\newcommand{\tabincell}[2]{\begin{tabular}{@{}#1@{}}#2\end{tabular}}
	\renewcommand\arraystretch{1.3}
	\scriptsize
	\centering
	\caption{Distribution of the galaxies in LAMOST DR4.}
	\label{tab:count_spec}
	\begin{tabular}{p{1.8cm}p{0.5cm}p{0.6cm}p{1.0cm}p{1.2cm}p{1.1cm}}
		\hline
		 & & Total & Region & Region & Region\\
		 & & & $b\geq$ $30^{\circ}$ & $b\leq$ $-30^{\circ}$ &$|b|<30^{\circ}$\\
		\hline
		\multirow{3}{*}{\centering Galaxies in DR4} & count & 118,506 & 77,154 & 32,453 & 8,899 \\
		 & percent & 100\% &65.1\% & 27.4\% & 7.5\%\\
		 & unique  & 110,154 & & & \\
		\hline
		\multirow{3}{*}{\tabincell{c}{Newly observed\\ galaxies}} & count & 40,182 & 11,472 & 24,347 & 4,363 \\
		 & percent & 100\% & 28.6\% & 60.6\% & 10.8\%\\
		 & unique  & 36,601 & & & \\
		\hline
	\end{tabular}
\end{table}

\subsection{Sample selection}

For the galaxies in LAMOST DR4, about 65\% spectra have spectral counterparts in SDSS DR13, while there are 40,182 spectra of 36,601 targets which have photometrical data but no spectroscopic observations in SDSS DR13. We select these newly observed galaxy spectra by LAMOST as our analytical sample in our work, called `Main' sample. In Figure \ref{fig:figure1}, the red points mark the footprints of these spectra. The number of these spectra are detailed in Table \ref{tab:count_spec}. We classify these galaxies to archive them in a catalogue, and calculate the composite spectra for different classes comparing with ones from SDSS spectra.

The target selection of LAMOST galaxies is unique in both spatial distribution and specific science aims. The geneal principle is to observe targets that SDSS fibers did not visit. Thus a great many objects in LAMOST input catalogue locate in Southern Galactic Cap region, and the objects in Northern Galactic Cap region are rejected by SDSS due to target density.

In the Northern Galactic Cap region of our input catalogue, there are some targets with highest priority in the area ( $-10^{\circ}$ $<\delta<$ $60^{\circ}$ and $b>$ $0^{\circ}$), where all the galaxies with $ r $-band Petrosian magnitude (Galactic reddening corrected) are brighter than r = 17.77 \citep{Shen et al2016}. These targets are candidates of galaxy pairs. A galaxy pair is typically defined from the projected distance $ r_p $ and recessional velocity difference |$ \Delta V $| of two neighbouring galaxies, and it is useful to probe the process of galaxy interactions or galaxy mergers(\citealt{Barton et al2000}, \citealt{Nikolic et al2004}). A large number of galaxy pairs are identified from SDSS main galaxy sample (\citealt{Ellison et al2008, Ellison et al2011}). However, a small fraction (< 10\%) of the SDSS main galaxy sample has not been targeted with spectroscopy due to the effect of fiber collisions. These missed galaxies have a very high probability of being in galaxy pairs. In order to obtain more galaxy pair candidates, these missed galaxies have been compiled into the input catalogue of LAMOST. In our Main sample, 2,859 spectra are from the input catalogue targeted of the SDSS missed candidates of galaxy pairs.

In the Southern Galactic Cap region($b\leq$ $-30^{\circ}$), there is a special survey strategy: The LAMOST Complete Spectroscopic Survey of Pointing Area(LCSSPA)\citep{Lam et al2015}, which is designed to complete the spectroscopic observations of all Galactic and extra-galactic sources which are selected from SDSS Data Release Nine, DR9(\citealt{Ahn et al2012}) using $ r $-band psf magnitudes and Petrosian magnitudes between 14.0 < r < 18.1 respectively in two selected fields of 20 square degrees. The central coordinates of the fields are ($ \alpha $, $ \delta $) = (37.88150939$ ^\circ $, 3.43934500$ ^\circ $) and (21.525988792$ ^\circ $, -2.200949833$ ^\circ $), respectively. In our Main sample, there are 4,493 galaxy spectra targeted from LCSSPA input catalogue. The scientific studies of the observed galaxies include galaxies, clusters of galaxies, and luminous infrared galaxies etc.

\subsection{Properties of the Main Sample}

\begin{figure*}
	\includegraphics[width=5.5cm]{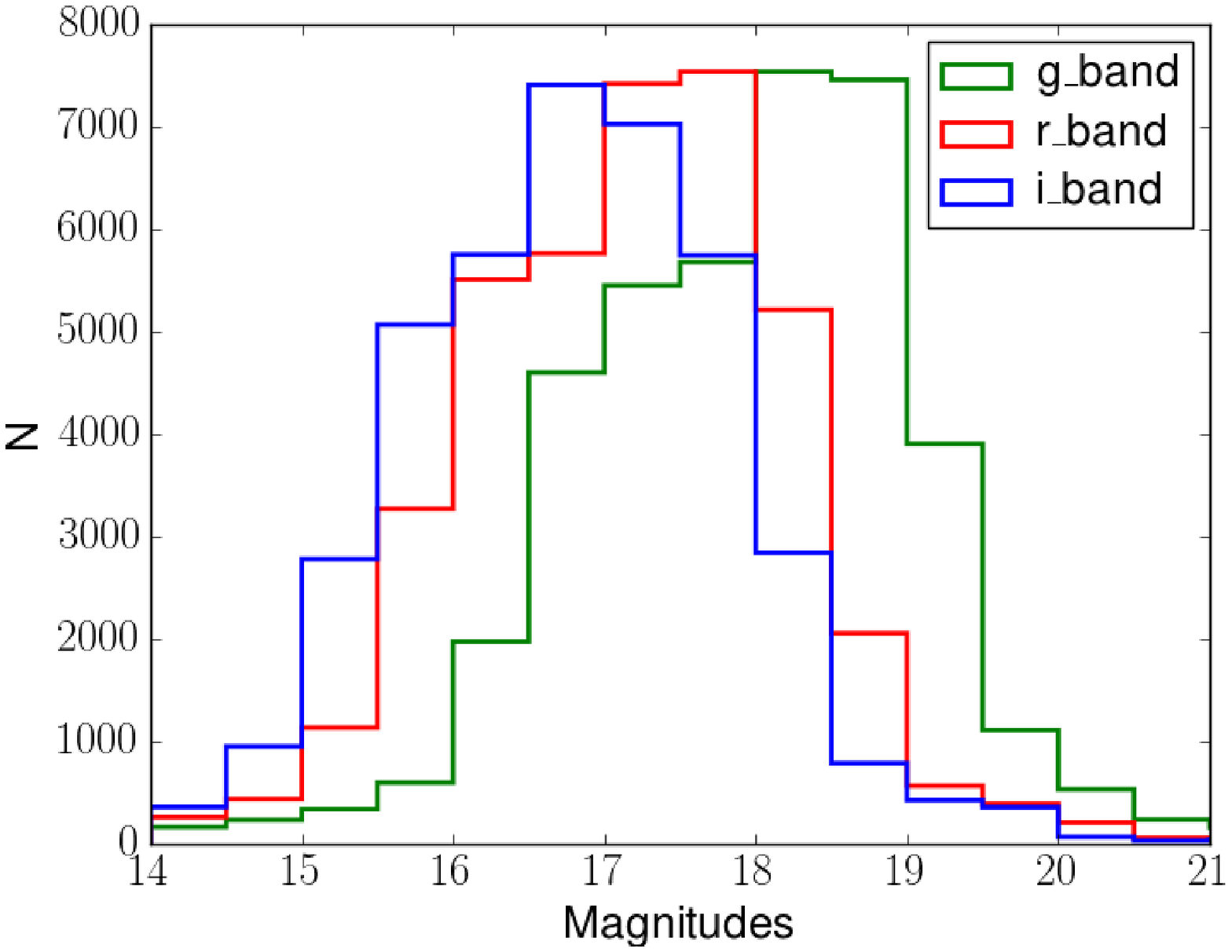}
	\includegraphics[width=5.5cm]{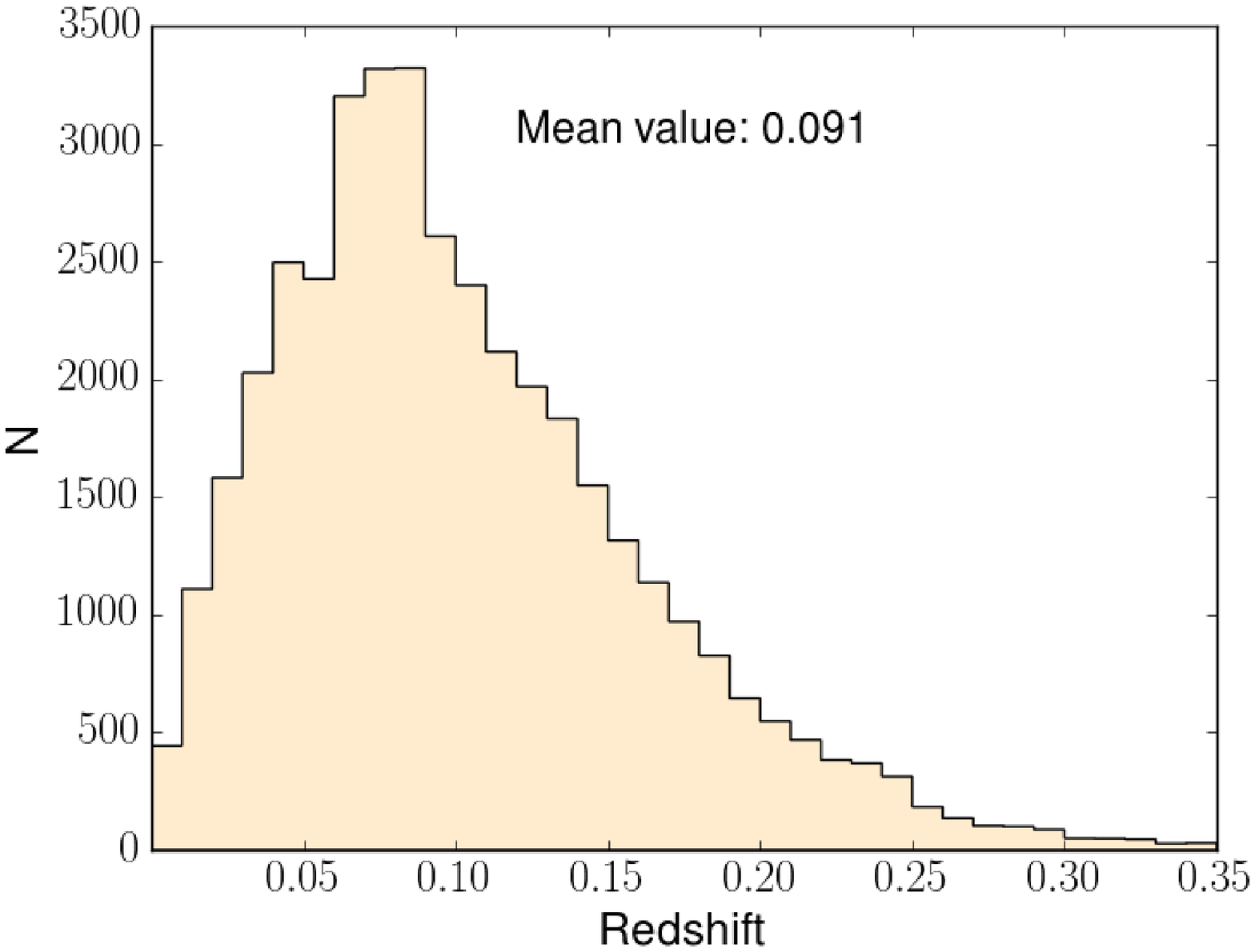}
	\includegraphics[width=5.5cm]{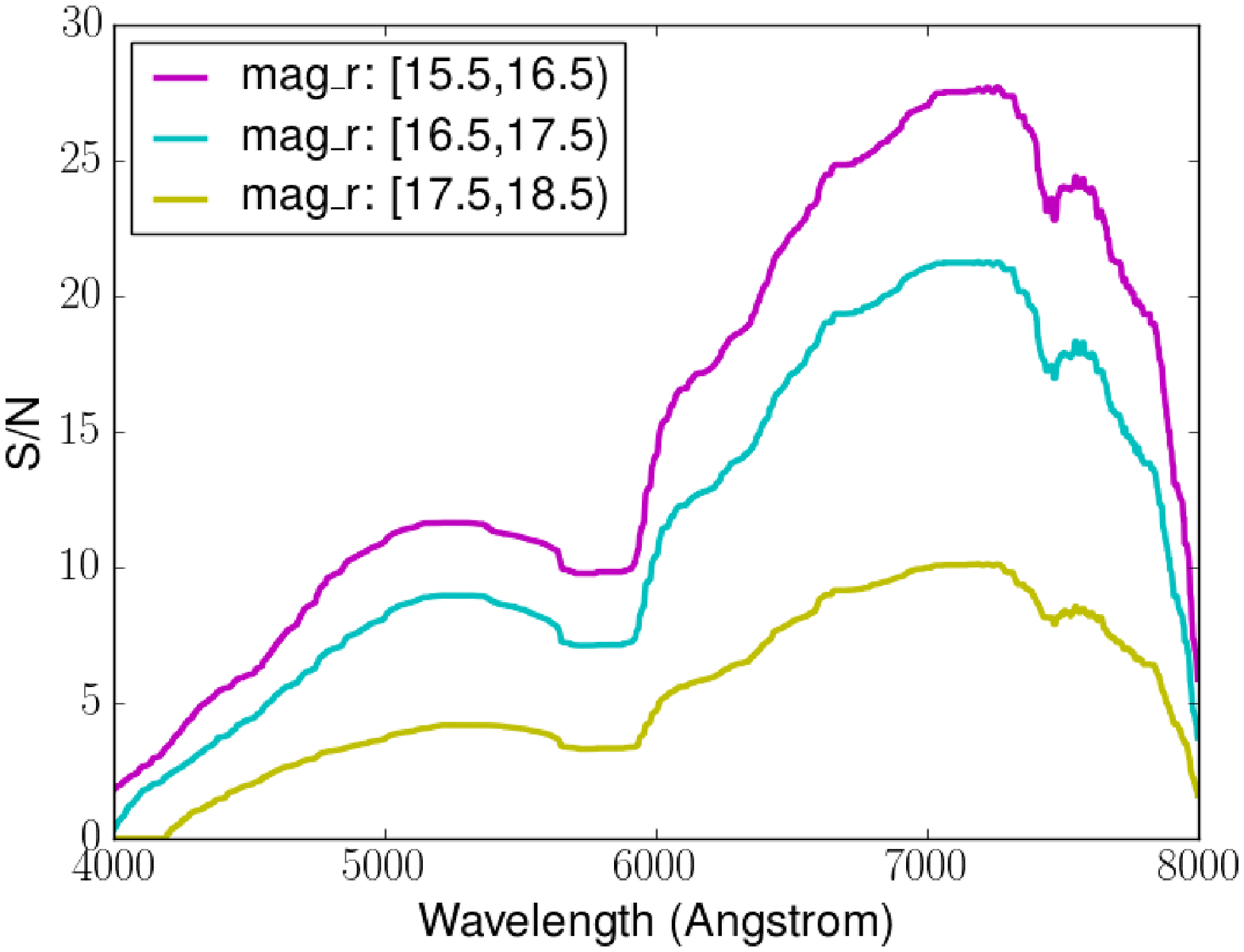}
    \caption{The left panel is the Petrosian magnitude distribution of $g$, $r$ and $i$ band, and the histograms with different colors represent the distribution of different bands: blue-- $g$ band, red-- $r$ band, and green-- $i$ band. The middle panel is the redshift distribution, and the mean value is 0.091. In the right panel, the curves represent the variation of S/N along with wavelength for magnitude bins in $r$ band: magenta color--[15.5,16.5), cyan color--[16.5,17.5), yellow color--[17.5,18.5), and the median values of S/N in $r$ band for the three magnitude ranges are 17, 13, 7 respectively.}
    \label{fig:mag_z_snr}
   \end{figure*}

Figure \ref{fig:mag_z_snr} displays the distribution features of our Main sample, including distributions of Petrosian magnitudes in $g$, $r$ and $i$ band, redshift and signal-to-noise ratio as a function of wavelength. The median values of magnitudes in $g$, $r$ and $i$ band are 18.0$^{mag}$, 17.2$^{mag}$, and 16.7$^{mag}$ respectively. The mean redshift is 0.091 and about 56.9\% of galaxies have redshift z $\leq$ 0.091, which is caused by target selection of LAMOST observations. The third panel presents the variation of S/N along with wavelength for three magnitude bins in $r$ band: [15.5,16.5), [16.5,17.5), [17.5,18.5). The median values of S/N in $r$ band for the three magnitude ranges are 17, 13, 7 respectively.

\section{Line intensity measurements}
\label{sec:line_m}

Precise line intensity measurement plays the key role in the following classification since the method we use is based on spectral line features. As is well known, galaxies display a very rich stellar absorption-line spectrum. Although late-type galaxies tend to have stronger emission lines and spectra dominated by hotter, more featureless stars, stellar Balmer absorption can still be substantial( 2-4\AA \ at H$ \beta $) \citep{Kauffmann et al2003b}. So the emission lines should be measured after the stellar absorption features subtracted. To address this problem, we can use population synthesis models by \cite{Bruzual et al2003}, hereafter BC03, to fit the continuum using a non-negative linear-least-squares routine. The fitting procedure automatically accounts for weak metal absorption under the forbidden lines and for Balmer absorption \citep{Bruzual et al2003}.

\begin{figure}
	\centering
	\includegraphics[width=8.0cm]{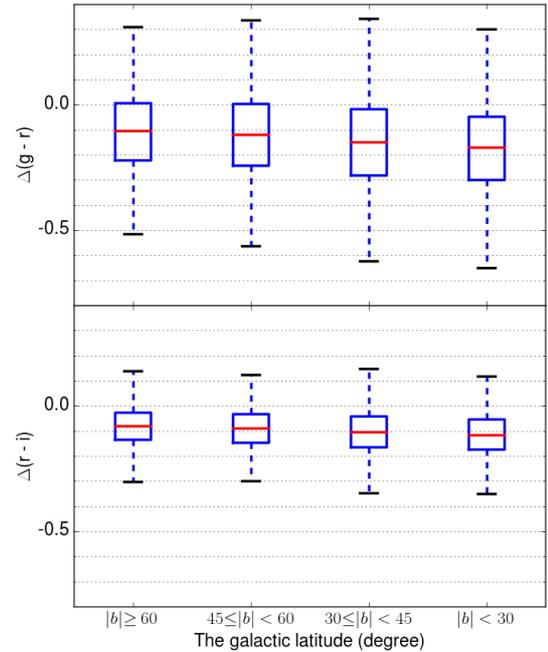}
    \caption{Differences between the synthetic colors $g-r$  and $r-i$ of some LAMOST spectra with their photometric counterparts of SDSS DR13 in galactic latitude bins (degree). The median color difference in high latitude($ |b|\geq60^{\circ} $) is found to be $\Delta (g-r) \approx$ -0.1 and $\Delta (r-i) \approx$ -0.08. And the lower latitude the spectra lie in, the larger difference they have (in low latitude($ |b|<30^{\circ} $) the median $\Delta (g-r) \approx$ -0.16 and $\Delta (r-i) \approx$ -0.11).}
    \label{fig:compare_beforcalib}
\end{figure}

Population synthesis methods require accurate continua of galaxy spectra. However, there are some uncertainties in the shape of the continua of LAMOST spectra caused by the current method of relative flux calibration \citep{Luo et al2015}, which may lead to an unprecise result of continuum-fitting. For the current used flux calibration of LAMOST spectra, some F type dwarfs with high quality spectra are chosen as standards to get the total response curve and the reddening of the standard stars is uncertain. So the extinction uncertainty of the standard stars might induce errors to the derived response curve, and thus leads to some uncertainty in flux calibration. And then it may bias the continuum shapes of the resulting spectra, especially severely effecting the spectra in areas with low galactic latitude. A check on the uncertainty of flux calibration is made by comparing synthetic magnitudes of LAMOST spectra with SDSS photometric magnitudes of the $ g $, $ r $, $ i $ bands( after correcting the Galactic extinction \citep{Schlegel et al1998}). Figure \ref{fig:compare_beforcalib} shows the color differences between them along with the galactic latitudes. The median color difference in high latitude($ |b|\geq60^{\circ} $) is found to be $\Delta (g-r) \approx$ -0.1 and $\Delta (r-i) \approx$ -0.08, and in low latitude($ |b|<30^{\circ} $) the median $\Delta (g-r) \approx$ -0.16 and $\Delta (r-i) \approx$ -0.11. The differences in both colors suggests that LAMOST spectra are bluer than SDSS photometry, and as the galactic latitude is lower, the difference is more and more greater. The most probable cause is that the uncertainty of reddening value of the standard stars contributes to some uncertainty in response curves\citep{Du et al2016}. And the lower latitude the spectra lie in, the more uncertainty the response curves have. So in order to obtain more accurate line intensity measurements, we re-calibrate the spectra of galaxies in LAMOST to correct the shape of continua of spectra.

\subsection{Re-calibration of LAMOST spectra}

\begin{figure}
   \centering
   \includegraphics[width=8.0cm]{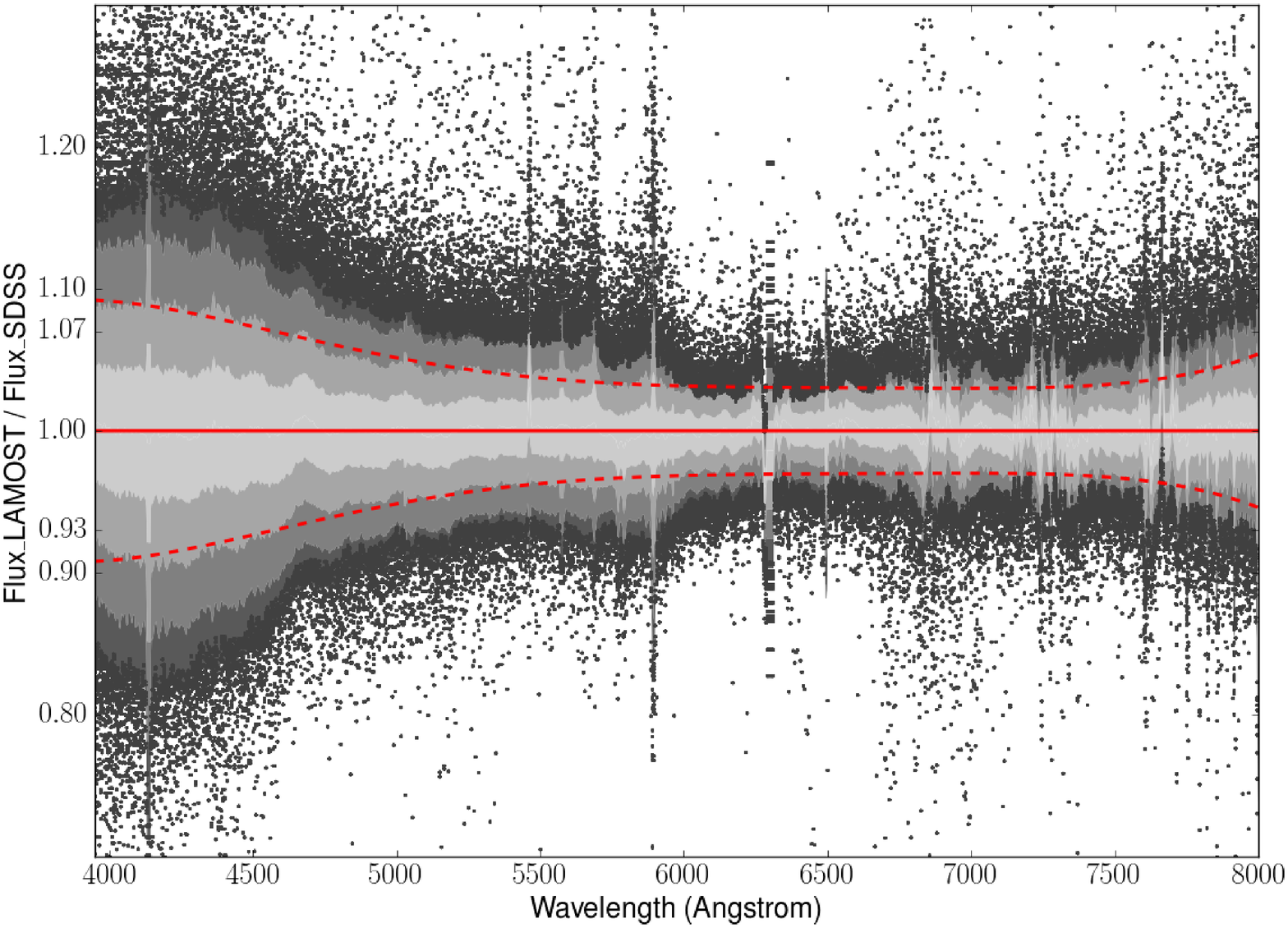}
   \includegraphics[width=8.8cm]{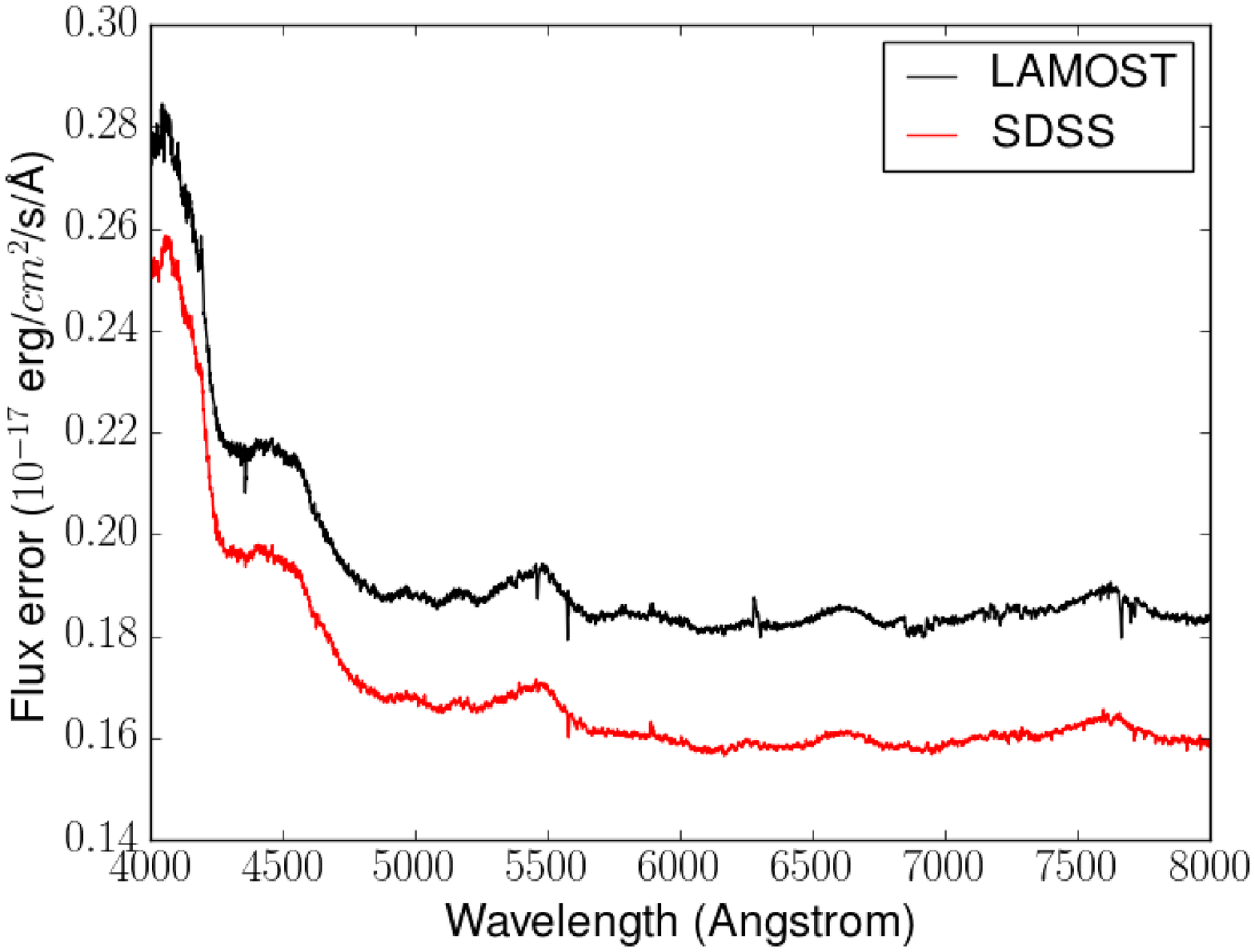}
   \caption{Comparison between LAMOST re-calibrated spectra and the spectra of the same sources from SDSS DR13 in a sample of 5000. On the upper panel, each point is a ratio value of LAMOST calibrated flux and corresponding SDSS one, the contour represents the distribution of the ratio values. The red solid and dash curves are the smoothed mean and standard deviation of the ratios, as a function of wavelength, respectively. On the lower panel, it is a comparison of median flux errors of LAMOST and SDSS, which shows LAMOST flux errors contribute more to the flux ratio.}
   \label{fig:compare_calib}
\end{figure}

We re-calibrate LAMOST fluxes using the $g$, $r$, and $i$ fiber magnitudes by cross-matching our sample with SDSS DR13 photometric catalogue. The original fluxes of LAMOST released spectra are corrected to the fluxes which are corresponding to photometric magnitudes by a low-order polynomial. To quantify the accuracy of our re-calibration, we calculate the difference between randomly selected 5,000 re-calibrated spectra of LAMOST and their SDSS spectral counterparts in observation wavelength. Before comparison, both spectra of LAMOST and SDSS are normalized by the median flux in the  4600-4800\AA \ region that is chosen to be devoid of strong emission lines, and rebinned to 1\AA. The difference is measured by the ratio between fluxes of LAMOST spectra and corresponding SDSS ones. In Figure \ref{fig:compare_calib}, the upper panel shows the comparison of the fluxes of LAMOST and SDSS, and the lower panel displays the comparison of the median flux error of LAMOST and SDSS.
On the upper panel, we can see the mean ratios (red solid line) are around 1.0 for the whole spectral wavelength coverage except for some region of sky emission lines and telluric bands attributed to the uncertainties of flat-fielding and sky-subtraction. The standard deviation (red dash curve) is less than 7\% in the wavelength range from 4,500 \AA \ to 8,000 \AA \ , and increases to 9\% in the band below 4,500 \AA \ due to the rapid decline of the instrumental throughput.
On the lower panel, the median errors in each wavelength bin of LAMOST flux are greater than those of SDSS, showing that LAMOST flux errors contribute more to the flux ratio.

\subsection{Line measurements}
\label{sec:line_m_process}

After the re-calibration, we correct the Galactic extinction using the reddening maps of \cite{Schlegel et al1998} and the extinction law of \cite{Cardelli et al1989}. And then the spectra are shifted into the rest frame.

As to measuring the line intensity, we use the method described as follows.

(i) Fit the stellar absorption features and continua of the galaxies. We adopt the basic assumption that any galaxy star formation history can be approximated as a sum of discrete bursts. So the spectrum of a galaxy is supposed to be a linear combination of individual stellar spectra of various types taken from a comprehensive library. Here we use a stellar population synthesis program called STARLIGHT \citep{Cid Fernandes et al2005}, which includes a library of template spectra with different ages and metallicities from the evolutionary synthesis models of BC03. We use three sets of templates, which host three metallicities(0.5, 1, 2.5 $Z_\odot$). Each set includes 11 different ages(0.005, 0.025, 0.1, 0.2, 0.6, 0.9, 1.4, 2.5, 5, 10, 13 Gyr). We fit galaxies with templates of single-metallicity populations (different ages) and obtain a best-fitting model spectrum that yields the minimum $ \chi^2 $.

(ii) Subtract the best-fitting stellar population model. The subtracted spectrum consists of three components\citep{Beck et al2016}: the emission lines, the noise and a slowly changing background that originates from the imperfect models. Since the emission lines and noise are high-frequency components, the background can be easily eliminated by a high-pass filter. We remove the possible retained background with a sliding 200 pixel median.

(iii) Fit the lines H$\beta$, [OIII]$\lambda$5007, H$\alpha$ and [NII]$\lambda$6585 on the residual spectrum of step (ii), which are mainly used in classification. Gaussians are used to simultaneously fit lines with automatically adjusting of the centers and widths to avoid deviation caused by redshift measurement. H$\beta$, [OIII]$\lambda$5007 are fitted with single Gaussian respectively. The H$\alpha$ and [NII]$\lambda$6585 are fitted with three Gaussians because the three lines [NII]$\lambda$6549, H$\alpha$ and [NII]$\lambda$6585 are blended in some galaxies, especially in LINER and Seyfert galaxies. Extensive visual inspection suggests that our line fitting method works well.

\begin{figure}
\centering
	\includegraphics[width=7.0cm]{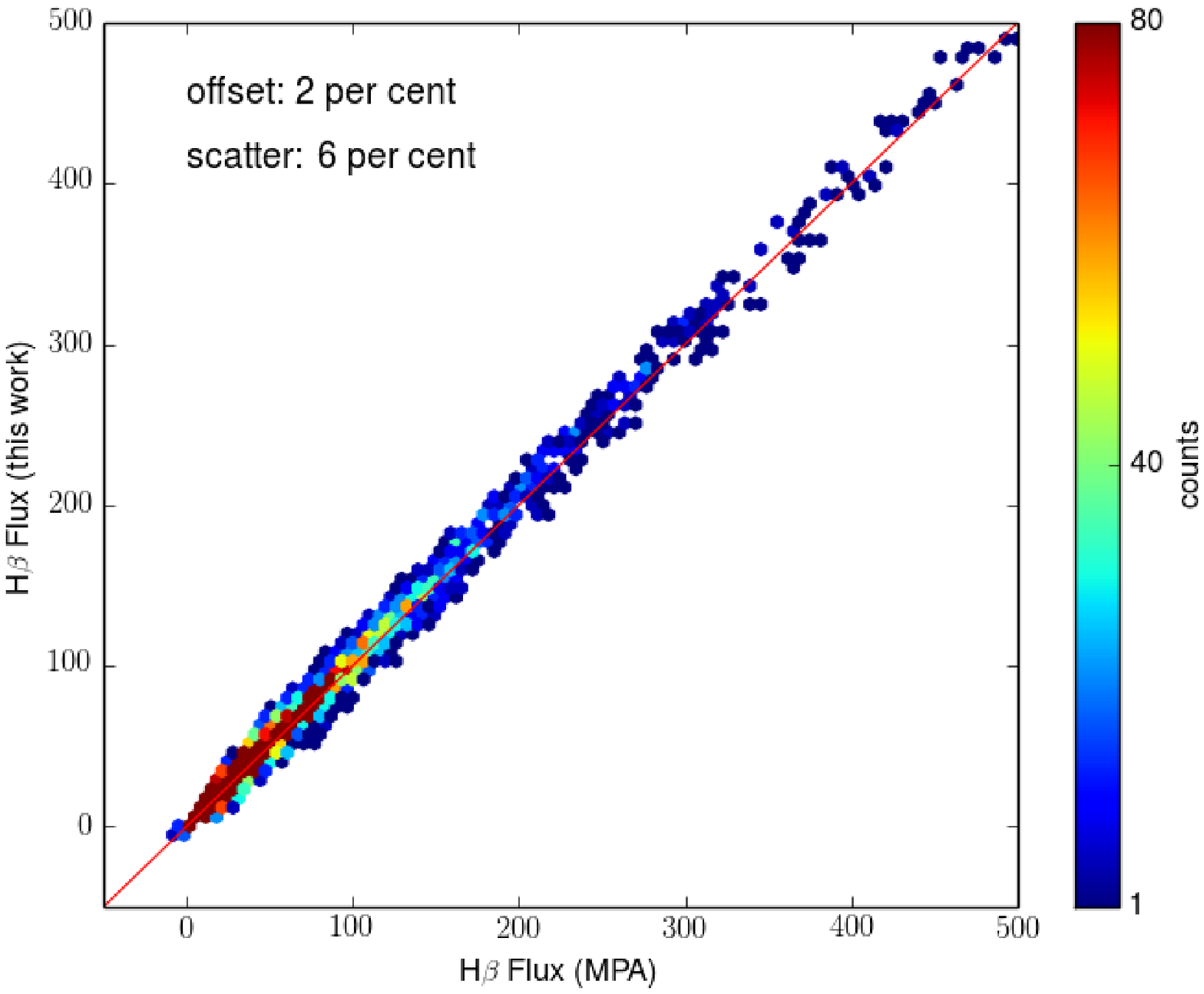}
	\includegraphics[width=7.0cm]{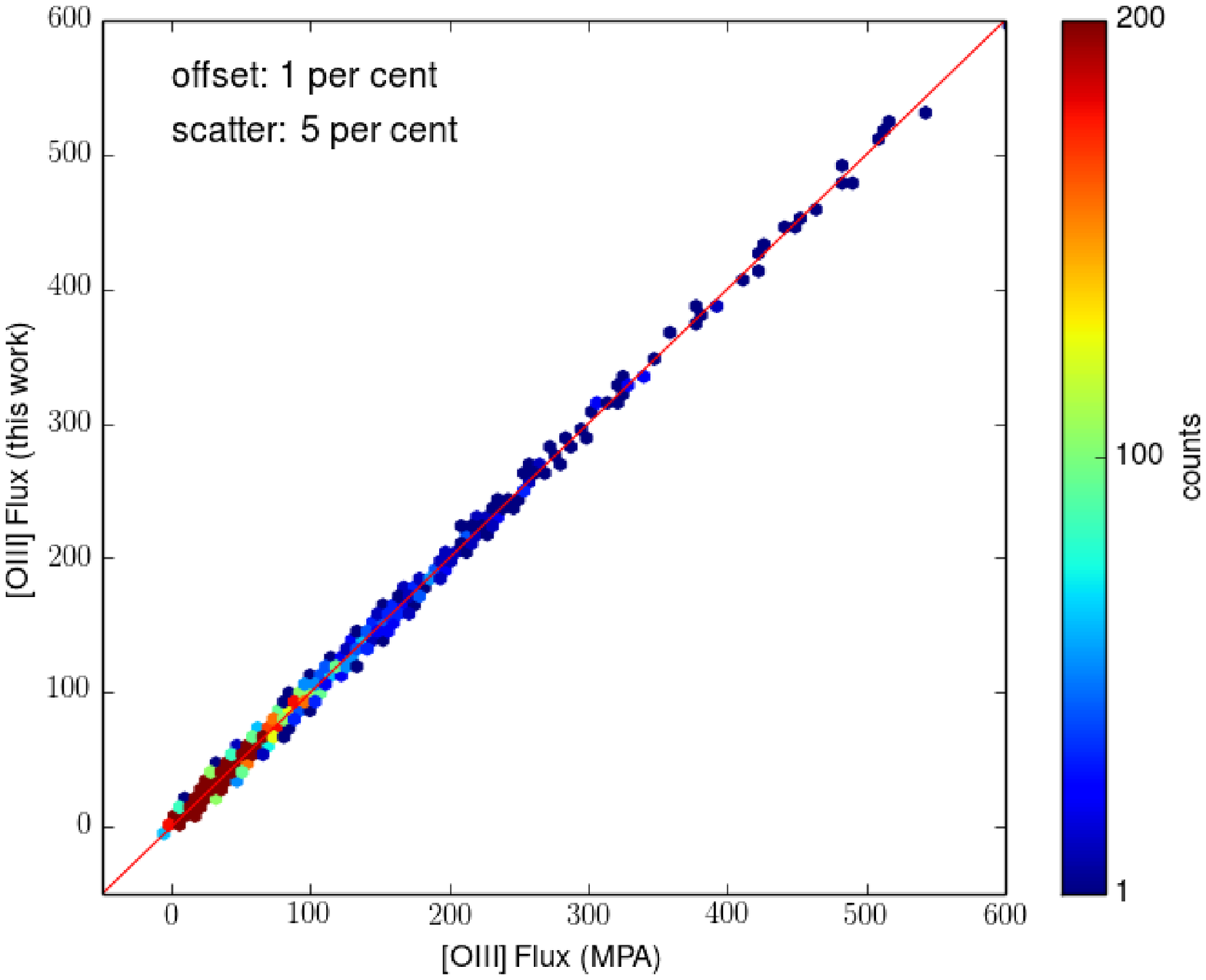}
	\includegraphics[width=7.0cm]{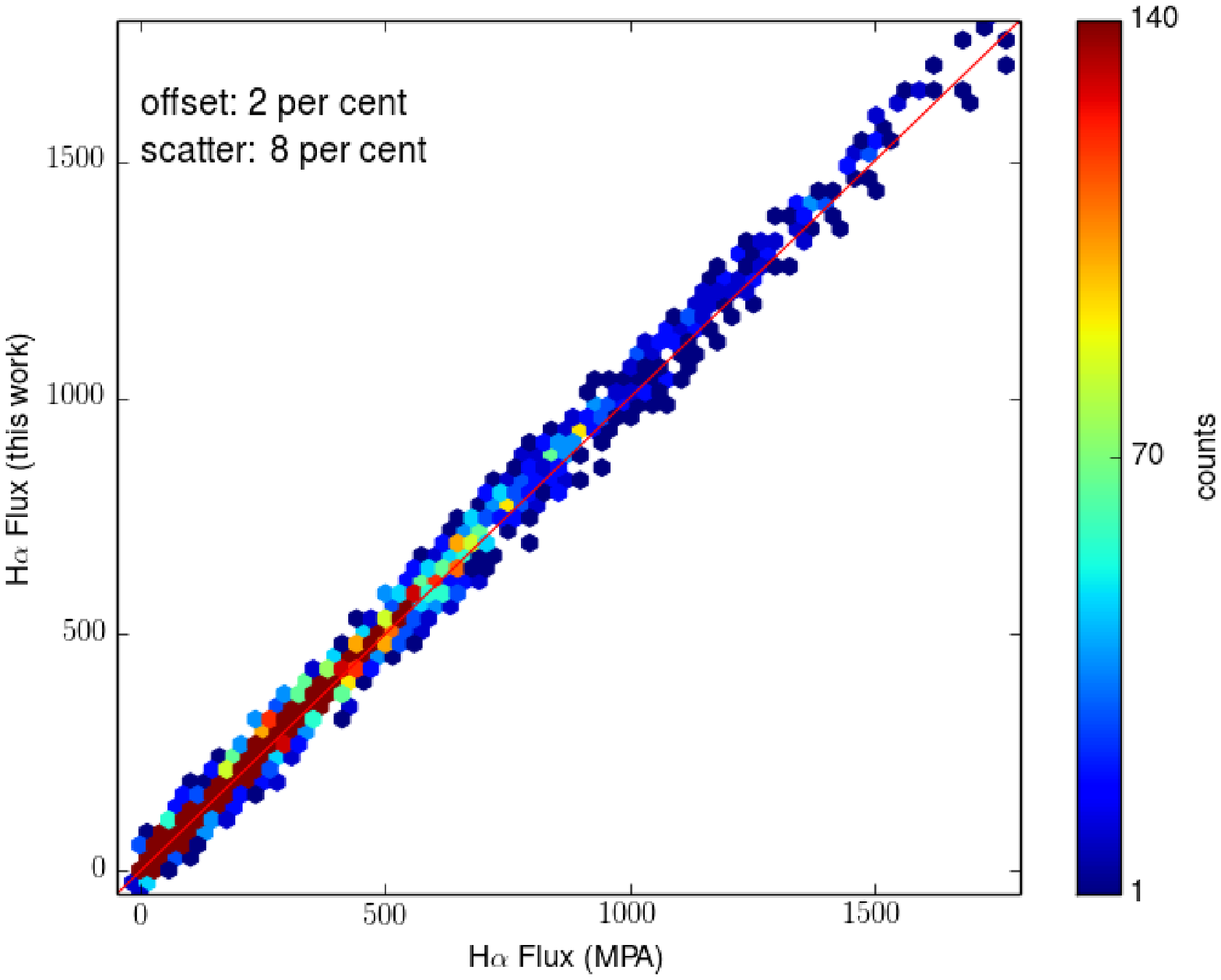}
	\includegraphics[width=7.0cm]{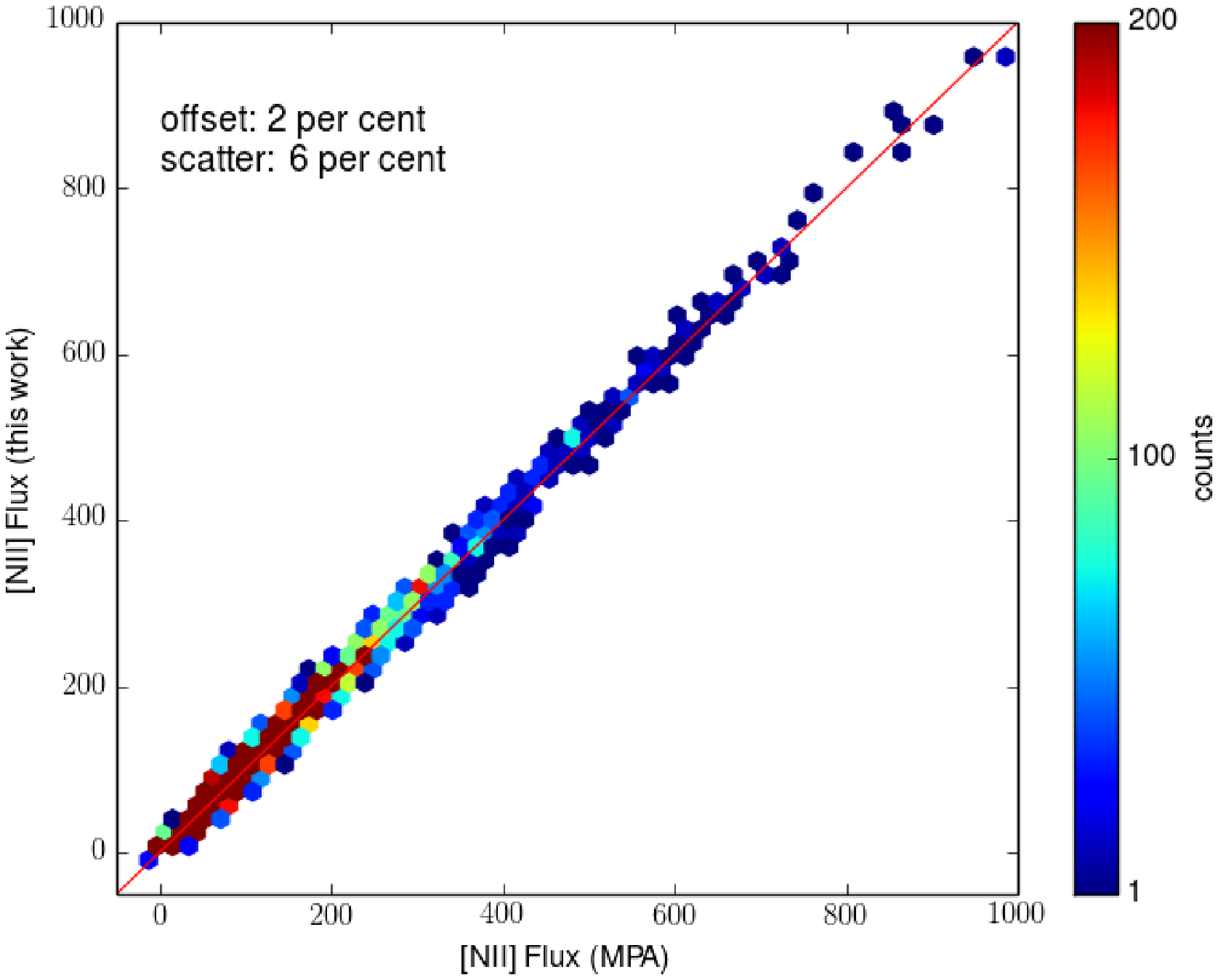}
    \caption{Comparison of the fluxes of four emission lines(H$\beta$, [OIII]$\lambda$5007, H$\alpha$ and [NII]$\lambda$6585) estimated by our method with ones presented by the MPA/JHU group. The red solid line is the identity line (y = x). The median offset and scatter are given on the left top of each panel. There is good agreements between the measurements. The largest dispersion is found for the fluxes of H$\alpha$, which is about 8 per cent. The colour indicates number of galaxies (scale given by the colour bar on the right-hand side).}
    \label{fig:flux_em}
\end{figure}

The MPA/JHU group provides publicly catalogue for flux measurements of the nebular emission lines of SDSS galaxies. It adopts a similar approach \citep{Tremonti et al2004} as our method. To test the reliability of our method of line intensity measurements, we compare the fluxes of nebular emission lines from MPA/JHU catalogue with our own estimations for SDSS galaxies. Figure \ref{fig:flux_em} shows comparisons between the fluxes of H$\beta$, [OIII]$\lambda$5007, H$\alpha$ and [NII]$\lambda$6585 measured by our code and those obtained from the MPA/JHU catalogue. The differences of all the four lines suggest that there is good agreement. The fluxes measured by our method are slightly larger than those of MPA/JHU catalogue( the median offset is 1-2 per cent), and the largest scatter ($ \sim $8 per cent) is found for the fluxes of H$\alpha$, probably due to different stellar population models for the estimates of underlying stellar absorption.

\textbf{In addition, we compare the line intensities between LAMOST spectra reduced by our code and their SDSS counterparts obtained from the MPA/JHU catalogue. Table \ref{tab:relative_diff} shows the relative differences of the line intensities of H$\beta$, [OIII]$\lambda$5007, H$\alpha$ and [NII]$\lambda$6585. From the table we can see that there are no significant differences of line intensities between LAMOST and SDSS, and this means that the LAMOST measurements and reduction process could reproduce the SDSS results although there exist small differences.}

\begin{table}
	\renewcommand\arraystretch{1.5}
	\centering
	\caption{Differences of the line intensities of repeated observations from LAMOST and SDSS. ((Intensity$_{LAMOST}$-Intensity$_{SDSS}$)/Intensity$_{SDSS}$)}
	\label{tab:relative_diff}
	\begin{tabular}{cc}
		\hline
		 & Relative differences\\
		\hline
		H$\beta$ & 3\%$\pm$8\%\\
		OIII$\lambda$5007 & 2\%$\pm$8\%\\
        H$\alpha$ & 1.5\%$\pm$7\%\\
        NII$\lambda$6585 & 2\%$\pm$6\%\\
		\hline
	\end{tabular}
\end{table}

\section{Classification of galaxies}
\label{sec:Classification}

H$\alpha$ line is the best quantitative indicator of the star formation rate in galaxies (\citealt{Kennicutt1992}). Based on H$\alpha$ together with other three lines: H$\beta$, [OIII]$\lambda$5007 and [NII]$\lambda$6585, we classify the galaxies into six classes: passive, H$\alpha$-weak, star-forming, composite, LINER and Seyfert, similar to the work of \cite{Dobos et al2012}, but with different boundaries of classes. The passive, H$\alpha$-weak galaxies are galaxies with no detectable emission lines or weak H$\alpha$ emission, while the other classes are emission-line galaxies.

\subsection{Classification schema}

First, we separate emission-line galaxies by the criterion that H$\alpha$ emission is detected at greater than 3 $ \sigma $. For these emission-line galaxies, we use BPT diagram to classify them based on H$\beta$, [OIII]$\lambda$5007, H$\alpha$ and [NII]$\lambda$6585. Note that we only use $\rm S/N_{H\alpha}$ $\geq$3 as the cut instead of $\rm S/N_\lambda$ $\geq$3 for all the four lines. In \cite{Cid Fernandes et al2010}, they presented if they adopt a uniform cut $\rm S/N_\lambda$ $\geq$3 for all the four lines, many weak line galaxies that have H$\beta$ and/or [OIII] below the threshold($\rm S/N_\lambda$ $\geq$3) are ignored in BPT diagram. So only the $\rm S/N_\lambda$ cut for H$\alpha$ is kept in our classification schema.

The emission-line galaxies are divided into star-forming galaxies, composite galaxies, LINERs and Seyferts based on the BPT diagram. There are several widely used empirical segregation curves on $\textsc{BPT}$ diagrams for classification, such as K01, K03 and K06. We use K03 and K01 as the border lines of star-forming galaxies and AGNs.

The empirical segregation curve of K03 is given in Equation~(\ref{eq:K03}). The star-forming galaxies locate below this curve.
\begin{equation}
    \rm log_{10}([OIII]/H\beta)=0.61/[log_{10}([NII]/H\alpha)-0.05]+1.3.
	\label{eq:K03}
\end{equation}

The empirical segregation curve of K01 is given in Equation~(\ref{eq:K01}). The pure-AGNs locate above this curve. The composite galaxies lie between K03 and K01.
\begin{equation}
    \rm log_{10}([OIII]/H\beta)=0.61/[log_{10}([NII]/H\alpha)-0.47]+1.19.
	\label{eq:K01}
\end{equation}

To distinguish Seyferts from LINERs, we use an alternative dividing line to K06. In K06, the emission-line ratios [OIII]/H$\beta$ versus [OI]/H$\alpha$ or [OIII]/H$\beta$ versus [SII]/H$\alpha$ are used to define Seyferts and LINERs. This means we should use extra lines [OI] or [SII] to classify Seyferts and LINERs besides H$\beta$, H$\alpha$, [OIII] and [NII], which might loss more emission-line galaxies\citep{Cid Fernandes et al2010}.  \cite{Cid Fernandes et al2010} transformed the K06 classification scheme into a simpler, more economic criterion (hereafter CF10). They still used H$\beta$, H$\alpha$, [OIII] and [NII] to classify Seyferts and LINERs without using extra emission lines. The border line is shown as Equation~(\ref{eq:CF10}).
\begin{equation}
    \rm log_{10}([OIII]/H\beta)=1.01*log_{10}([NII]/H\alpha)+0.48.
	\label{eq:CF10}
\end{equation}

In our work, we use Equation~(\ref{eq:CF10}) as the segregation curve to separate Seyferts from LINERs. This criterion is not only more economic than K06, but also avoid ambiguous classification in K06 involving more than one diagnostic diagrams \citep{Cid Fernandes et al2010}.

\begin{figure}
    \centering
	\includegraphics[width=8.5cm]{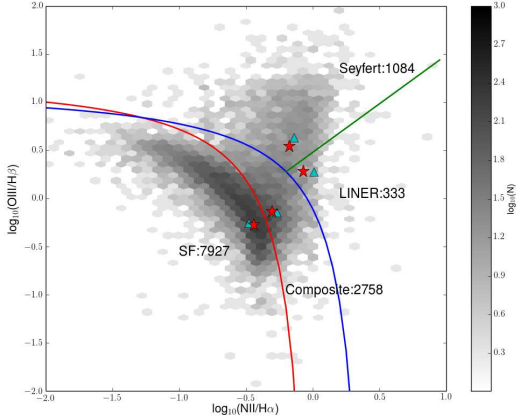}
    \caption{BPT diagram for emission-line galaxies in our samples. The density plot is the number density of galaxies in our Main sample. The red solid line is defined by K03, the blue dash line is defined by K01, and the green dot dash line is given by CF10. The numbers indicate the results of classification for star-forming galaxies, composite galaxies, LINERs and Seyferts. The red stars mark the loci of composite spectra of our Main sample described in Section \ref{sec:composite_em}. The cyan triangles are the loci of the composite spectra of SDSS, as seen in Section~\ref{sec:comparison_sec} in detail.}
    \label{fig:bpt_em}
\end{figure}

Emission-line galaxies in our Main sample are classified into star-forming galaxies, composite galaxies, LINERs and Seyferts using BPT diagram, shown in Figure \ref{fig:bpt_em}. The density plot is the number density of galaxies in our Main sample, and the three boundary lines are: the red solid line, blue dash line, and green dot dash line represent Equation~(\ref{eq:K03}),  (\ref{eq:K01}), and (\ref{eq:CF10}) respectively. The red stars mark the loci of composite spectra of our Main sample described in Section \ref{sec:composite_em}. The cyan triangles are the loci of the composite spectra of SDSS, as seen in Section~\ref{sec:comparison_sec} in detail.

Passive galaxies and H$\alpha$-weak galaxies are selected from all galaxies by $\rm S/N_{H\alpha}<$ 3. We consider the passive galaxies as galaxies with completely no H$\alpha$ emission. The H$\alpha$-weak galaxies have measurable, but weak H$\alpha$ emission.

In brief, we classify our sample into six classes.

(1) Passive galaxies: no measurable H$\alpha$ emission.

(2) H$\alpha$-weak: H$\alpha$ emission is measurable, but weak, not strong enough for classification using the $\textsc{BPT}$ diagram.

(3) Star-forming galaxies: below the red line defined by Equation~(\ref{eq:K03}) in Figure \ref{fig:bpt_em}. Galaxies in this class are with evident star formation signatures but no active nucleus.

(4) Composite galaxies: between the red line defined by Equation~(\ref{eq:K03}) and the blue line defined by Equation~(\ref{eq:K01}) in Figure \ref{fig:bpt_em}. Spectra of this class are mixture of the features of $\textsc{AGN}$ and star formation galaxy.

(5) LINER galaxies: above the blue line defined by Equation~(\ref{eq:K01}) and below the green line defined by Equation~(\ref{eq:CF10}) in Figure \ref{fig:bpt_em}. Galaxies in this class are AGNs in low ionization nuclear emission-line regions.

(6) Seyfert galaxies: between the blue line defined by Equation~(\ref{eq:K01}) and the green line defined by Equation~(\ref{eq:CF10}) in Figure \ref{fig:bpt_em}. Galaxies in this class are Seyfert II with narrow emission lines.

\subsection{Classification results}

The number and percentage of each galaxy class obtained with the above classification scheme are shown in Table \ref{tab:class_count}. In our Main sample, the number of H$\alpha$-weak galaxies is the highest, while the number of LINER galaxies is the lowest. The distribution of percentage is a little different to the result of SDSS galaxies classes in \cite{Dobos et al2012}: the percentages of emission-line galaxies (star-forming galaxies, composite galaxies, LINERs and Seyferts) in our classification are higher than those of \cite{Dobos et al2012}, because we use the cut of emission-line galaxies only by $\rm S/N_\lambda$ $\geq$3 for H$\alpha$, while \cite{Dobos et al2012} uses $\rm S/N_\lambda$ $\geq$3 for the four lines. The ratio of Seyferts to LINERs is smaller than that in \cite{Dobos et al2012}, which may be caused by the different methods of Seyfert/LINER separation. In Dobos et al. (2012), they used LINER/Seyfert separation presented by K06: [OIII]/H$\beta$ versus [OI]/H$\alpha$. If we classify the AGNs in our sample to Seyferts and LINERs by the same method as \cite{Dobos et al2012}, the ratio of Seyferts to LINERs is simlar to \cite{Dobos et al2012}.

\begin{table}
	\renewcommand\arraystretch{1.5}
	\centering
	
	\caption{Statistical results of the classification.}
	\label{tab:class_count}
	\begin{tabular}{p{0.6cm}p{0.6cm}p{1.2cm}p{0.6cm}p{1.0cm}*{2}{p{0.6cm}}}
		\hline
		 & passive & H$\alpha$-weak & SF & Composite & LINER & Seyfert\\
		\hline
		count & 10579 & 17032 & 7927 & 2758 & 333 & 1084\\
		percent & 26.64\% & 42.88\% & 19.96\% & 6.95\% & 0.84\% & 2.73\%\\
		\hline
	\end{tabular}
\end{table}

\subsection{Correlation with morphological class}
Hubble type of a galaxy closely correlates with its spectrum (\citealt{Kennicutt1992}). In \cite{Kennicutt1992}, he provided spectra of a set of 55 nearby galaxies with known Hubble types. The set contained all Hubble types, from giant ellipticals to dwarf irregulars. He presented that the elliptical galaxies are dominated by absorption features and nebular emission lines are absent or weak, while in Sbc--Sc galaxy spectra the principal nebular emission lines are apparent with intensities which are characteristic of star-forming HII regions. In this section, we roughly correlate our spectral classes with morphological classes quantitatively.

There are some frequently used parameters sensitive to morphology, such as concentration index(\citealt{Shimasaku et al2001}, \citealt{Strateva et al2001}, \citealt{Park Choi2005}), color(\citealt{Morgan et al1957}, \citealt{Strateva et al2001}), and D4000$_n$ index (\citealt{Kauffmann et al2003c}, \citealt{Mateus et al2006}). The concentration index is defined to be the ratio of the radii containing 90 and 50 percent of the Petrosian flux in the $r$ band ($C = R90/R50$). \cite{Shimasaku et al2001} found a strong correlation between $C$ and morphological types and suggested $C$ $\sim$3 as a criterion of morphological classification of galaxies. However, they also noted that it is difficult to construct a pure early-type galaxy sample based only on the concentration index, since the resulting sample has $\sim$20 percent contamination by late-type galaxies. \cite{Strateva et al2001} have shown that there is a good correspondence between $C$ and Hubble type: elliptical galaxies have values around 3 and disc-dominated galaxies have values around 2--2.5, with $C$ $\sim$2.6 marking the boundary between early(elliptical and lenticular) and late types(spiral and irregular). They also analyzed the ($ u $ - $ r $) color which is a more conventional estimator of galaxy types, finding that ($ u $ - $ r $) color=2.22 clearly separated early and late morphological type. 4000 \AA \ break is another parameter that can be used to classify early and late type galaxies, which is small for galaxies with younger stellar populations, and large for older galaxies(\citealt{Kauffmann et al2003b}, \citealt{Mateus et al2006}). Here we compute the three indicators of spectra in different classes and analyze the correlation with spectral classes and the morphological types.

\begin{figure*}
	\includegraphics[width=5.5cm]{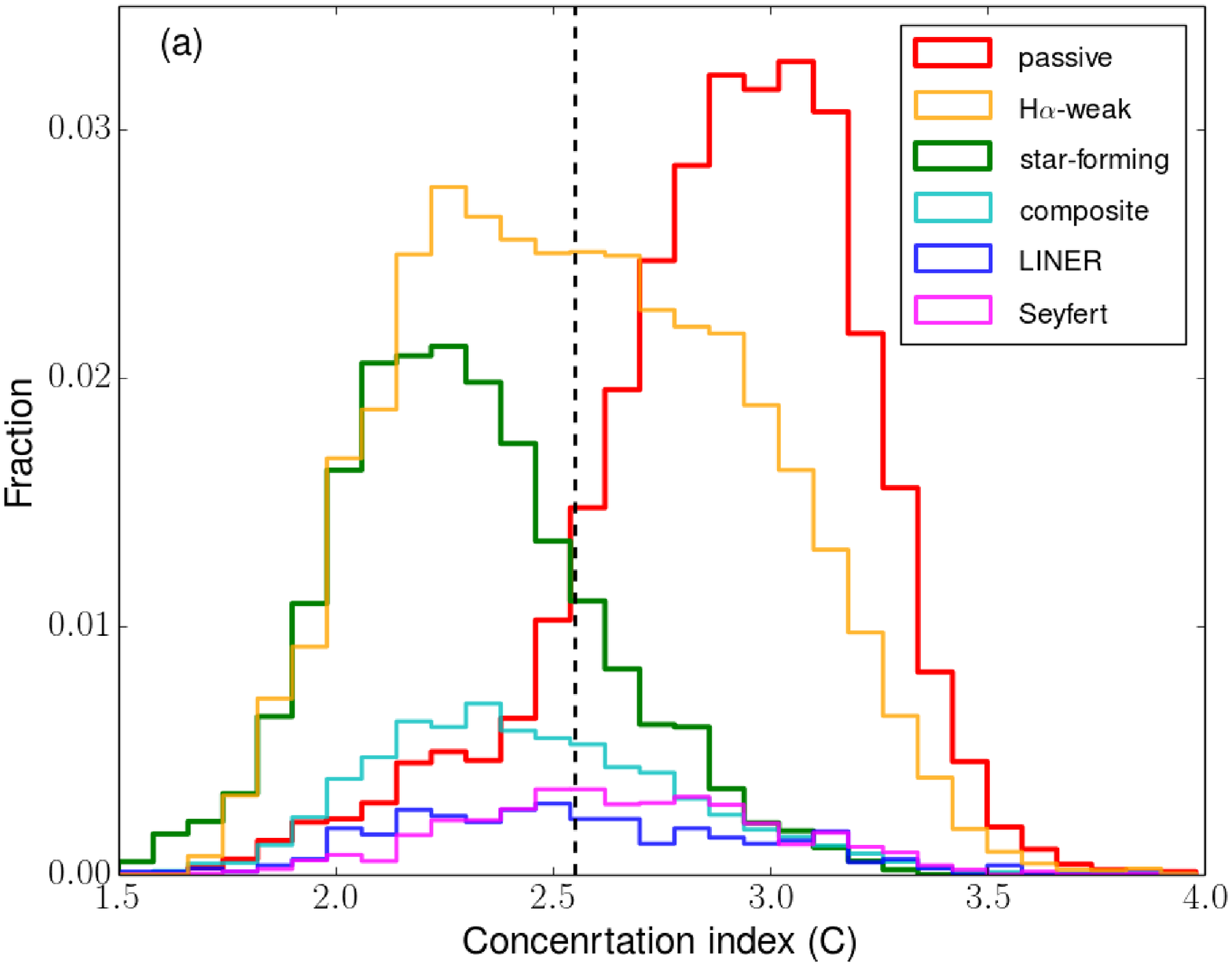}
	\includegraphics[width=5.5cm]{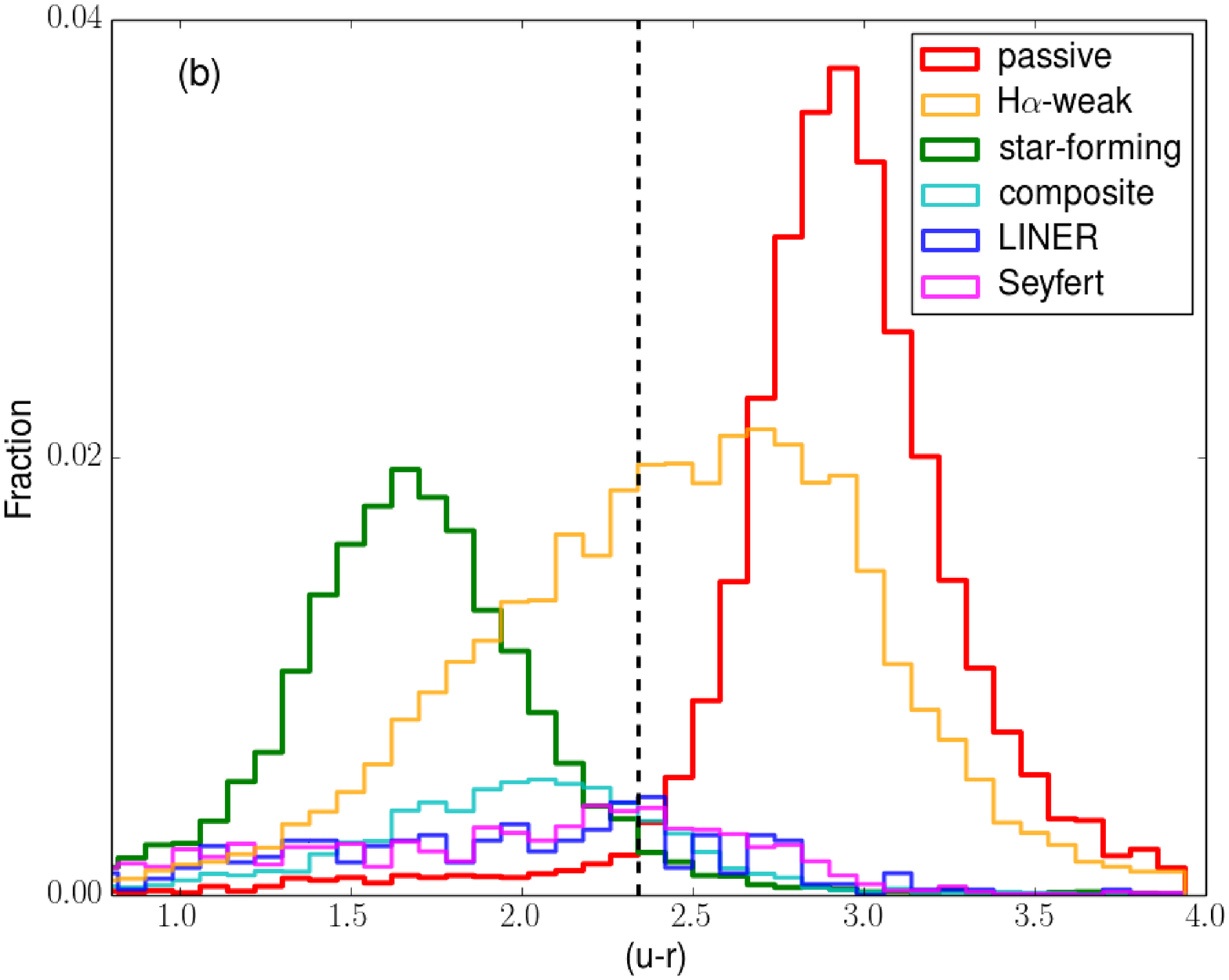}
	\includegraphics[width=5.5cm]{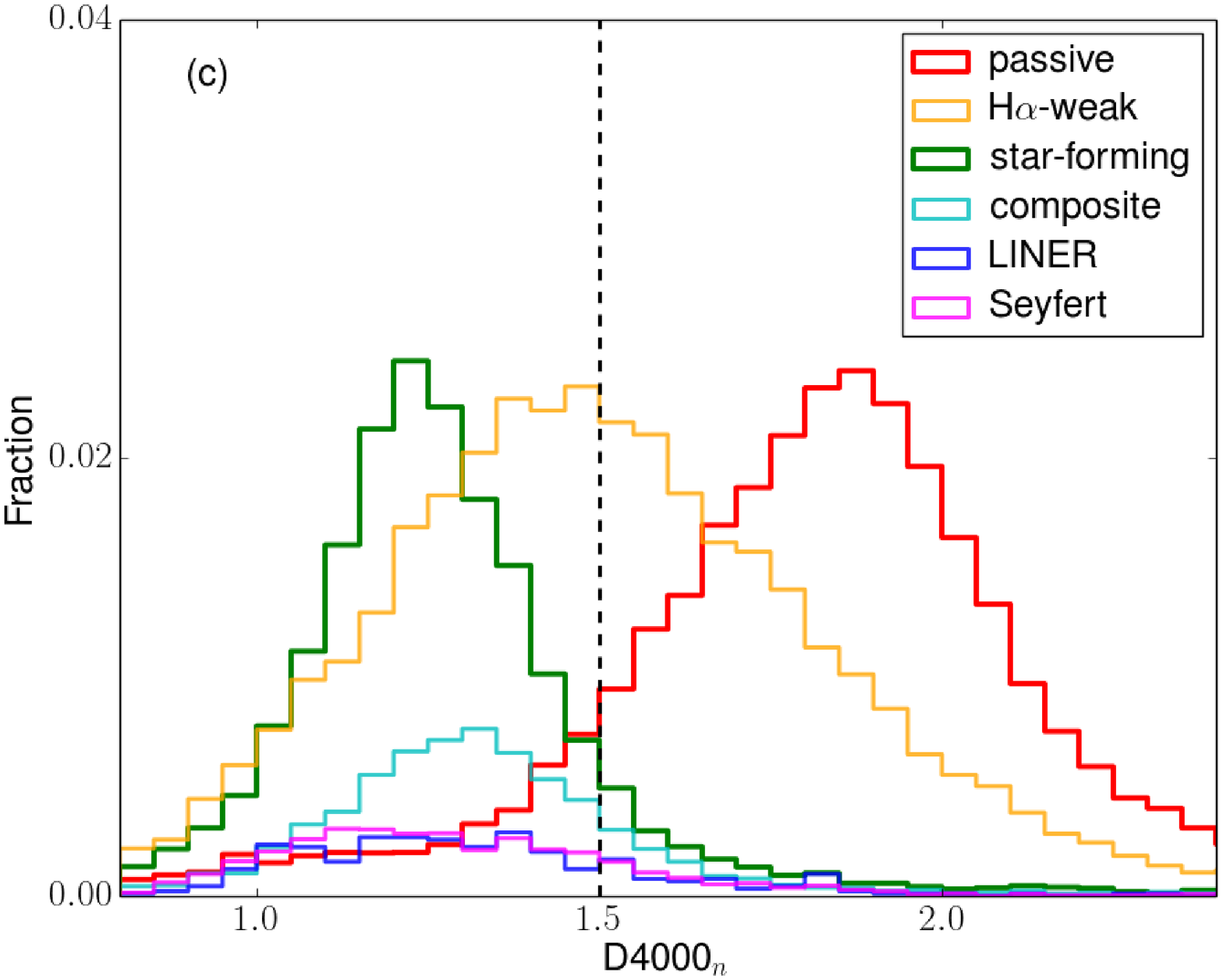}
    \caption{Distribution of concentration index(panel a), ($ u $ - $ r $) color(panel b), D4000$_n$ index(panel c) for each class in our galaxy sample. The vertical dashed lines are the best separators for star-forming and passive types. Note that in these three panels, the fractions of LINERs in each bins are enlarged 5 times in order to make the low-count curves of LINERs visible.}
    \label{fig:distribution_C}
\end{figure*}

We compute the concentration index of spectra in different classes using the values of petroR90 and petroR50 in the $r$ band which are cross-matched with the photometric catalogue of SDSS. The magnitudes of $ u $ and $ r $  band are also obtained from SDSS to compute the ($ u $ - $ r $) color. For the 4000\AA \ break, we use the narrow definition D4000$_n$ introduced by \cite{Balogh et al1999}, which is the ratio of the average flux density in the bands 3850-3950 \AA \ and 4000-4100 \AA. In Figure \ref{fig:distribution_C}, we show the distributions of concentration index, ($ u $ - $ r $) color, D4000$_n$ index for each spectral class in our galaxy sample: passive, H$\alpha$-weak, star-forming, composite, LINER and Seyfert. Table \ref{tab:medianvalues} lists the median values of these three parameters for different spectral classes. From Figure \ref{fig:distribution_C} and Table \ref{tab:medianvalues},  the star-forming galaxies are bluer, less concentrated and have smaller D4000$_n$ index; on the contrary, passive galaxies are redder, more concentrated and have larger D4000$_n$ index. We also note that the other four classes occupy an intermediate locus in color, concentration index and D4000$_n$ index distributions, showing the mix of morphological classes.

\begin{table}
	\renewcommand\arraystretch{1.3}
	\centering
	\caption{The median values of the three parameters: concentration index($ C $), ($ u $ - $ r $) color, D4000$_n$ index for different spectral classes.}
	\label{tab:medianvalues}
	\begin{tabular}{p{0.8cm}p{0.8cm}p{1.0cm}p{0.8cm}p{1.0cm}*{2}{p{0.8cm}}}
		\hline
		 & passive & H$\alpha$-weak & SF & composite & LINER & Seyfert\\
		\hline
		 $ C $ & 2.93 & 2.55 & 2.28 & 2.42 & 2.51 & 2.65\\
		 ($ u $-$ r $) & 2.91 & 2.48 & 1.66 & 1.99 & 1.98 & 1.97\\
		 D4000$_n$ & 1.85 & 1.49 & 1.25 & 1.31 & 1.29 & 1.25\\
		\hline
	\end{tabular}
\end{table}

In Figure \ref{fig:distribution_C}, the black dash line displays an optimal value that separates those two extreme classes: star-forming and passive galaxies. Following \cite{Mateus et al2006}, we define two parameters: reliability and completeness, and then we find the optimal value that maximizes the product of reliability and completeness. The reliability is the fraction of galaxies from a given spectral class that are correctly classified by using the optimal value, and the completeness is the fraction of all galaxies given spectral class that are actually selected. $ \mathcal{R}_{SF}$, $\mathcal{R}_{P}$, $\mathcal{C}_{SF}$ and $\mathcal{C}_{P} $ represent the reliability and completeness for star-forming galaxies and passive galaxies, respectively. We find that $ C $ = 2.55, ($ u $ - $ r $ ) = 2.3 and D4000$_n$=1.5 are the optimal separators among star-forming galaxies and passive galaxies, which are close to the values obtained by \cite{Mateus et al2006}. The reliability and completeness of these separators are shown as Table \ref{tab:separator}. In addition, these three boundary values are also close to the values given in previous literatures to separate early types and late types( \citealt{Strateva et al2001}, Kauffmann et al. 2003a,b).

\begin{table}
	\renewcommand\arraystretch{1.3}
	\centering
	\caption{Reliability and completeness for three separators of concentration index($ C $), ($ u $ - $ r $) color, D4000$_n$ index obtained to split the distributions of star-forming galaxies and passive galaxies.(in percent)}
	\label{tab:separator}
	\begin{tabular}{cccccc}
		\hline
		 & Separator & $ \mathcal{R}_{SF}$ & $\mathcal{C}_{SF}$ & $\mathcal{R}_{P}$ & $\mathcal{C}_{P} $\\
		\hline
		 $ C $ & 2.55 & 90.1 & 88.4 & 81.8 & 82.3\\
		 ($ u $ - $ r $) & 2.3 & 96.3 & 92.1 & 85.4 & 90.1\\
		 D4000$_n$ & 1.5 & 91.2 & 93.0 & 85.4 & 81.7\\
		\hline
	\end{tabular}
\end{table}

\subsection{Galaxy catalogue}
\label{sec:gal_catalog}

We present a catalogue of galaxies for our Main sample, which have photometric information without spectroscopic observations in SDSS. One aim of our work is to provide a catalogue for the newly observed galaxies in LAMOST, which have no spectral observations in SDSS. We give the fluxes of nebular emission lines and the classification information in this catalogue. There are 40,182 entries (36,601 targets) in the catalogue, and Table~\ref{tab:catalog_Main} lists a part of this catalogue. For each item in Table~\ref{tab:catalog_Main}, there are five kinds of information: basic information retrieved from the LAMOST catalogue; Petrosian magnitudes cross-matched with the photometric catalogue of SDSS; three parameters of correlation with morphological types; the fluxes of lines we computed in Section \ref{sec:line_m}; and spectral class using our classification scheme. Note that, the morphological types are not directly given, instead, three parameters related to the morphological types: concentration index, ($ u $ - $ r $) color, D4000$_n$ index are provided. Records are marked with `$ a $' at the upper right corner of the designation if the spectra are from a subsample in Northern Galactic Cap which includes 2,859 spectra complementary to SDSS main galaxy sample in the study of galaxy pairs. While records marked with `$ b $' represent the spectra targeted from The LAMOST Complete Spectroscopic Survey of Pointing Area at Southern Galactic Cap, including 4,493 spectra. The complete catalogue is available at http://sciwiki.lamost.org/downloads/wll.

\begin{table*}
   \renewcommand\arraystretch{1.4}
    %\tiny
	%\footnotesize
    \scriptsize
	    \caption{The galaxy catalogue with classification and related information for our Main sample.}
        \begin{tabular}{p{2.3cm}<{\raggedleft}p{0.9cm}<{\raggedleft}p{1.1cm}<{\raggedleft}p{0.5cm}<{\raggedleft}p{0.3cm}<{\raggedleft}p{0.6cm}<{\raggedleft}p{0.6cm}<{\raggedleft}p{0.6cm}<{\raggedleft}p{0.6cm}<{\raggedleft}p{0.6cm}<{\raggedleft}p{0.35cm}<{\raggedleft}p{0.35cm}<{\raggedleft}p{0.4cm}<{\raggedleft}p{0.7cm}<{\raggedleft}p{0.55cm}<{\raggedleft}p{0.55cm}<{\raggedleft}p{0.55cm}<{\raggedleft}p{0.4cm}<{\raggedleft}}
		\hline
		Designation & Ra & Dec & z & S/N\_r & Mag\_u & Mag\_g & Mag\_r & Mag\_i & Mag\_z & C & u-r & D4000$_n$ & H$\beta$\_flux & OIII\_flux & H$\alpha$\_flux & NII\_flux & class\\
		(1) & (2) & (3) & (4) & (5) & (6) & (7) & (8) & (9) & (10) & (11) & (12) & (13) & (14) & (15) & (16) & (17) & (18)\\
		\hline		
		J111054.26+241058.9$^a$ & 167.726087 & 24.183041 & 0.11794 & 7.46 & 20.3184 & 18.76986 & 17.7697 & 17.34902 & 17.05103 & 2.83 & 2.9 & 2.17 & -25.95 & -9.3 & 4.43 & -5.77 & 0 \\
J233728.25+062508.5 & 354.36771 & 6.41905 & 0.18097 & 6.81 & 20.85624 & 19.00901 & 17.82978 & 17.35194 & 17.03384 & 2.9 & 3.13 & 1.39 & -1.28 & 7.68 & -4.82 & 28.99 & 0 \\
J013101.46-014335.9 & 22.756086 & -1.726659 & 0.06835 & 3.04 & 25.0782 & 18.61482 & 18.05775 & 17.83633 & 18.12357 & 2.35 & 2.84 & 1.14 & 9.51 & 12.0 & 40.82 & 8.68 & 1 \\
J105708.55+191040.3 & 164.285643 & 19.177862 & 0.04511 & 16.59 & 18.69523 & 17.38144 & 16.90178 & 16.59306 & 16.36377 & 1.98 & 1.64 & 1.16 & 98.58 & 38.44 & 490.31 & 180.25 & 1 \\
J023527.55+015852.7$^b$ & 38.8648126 & 1.9813188 & 0.07355 & 20.73 & 18.62122 & 17.3105 & 16.79459 & 16.49822 & 16.32904 & 2.22 & 1.99 & 1.47 & 34.13 & 33.78 & 210.29 & 93.20 & 2 \\
J024553.28+011830.0 & 41.47202 & 1.30835 & 0.13325 & 1.61 & 19.97601 & 18.9666 & 18.35996 & 18.03479 & 17.741 & 2.38 & 1.97 & 1.24 & 36.09 & 79.59 & 143.3 & 72.44 & 2 \\
J160958.51+460950.2 & 242.493808 & 46.163967 & 0.08866 & 32.28 & 19.00142 & 17.56057 & 16.83426 & 16.39977 & 16.17576 & 3.04 & 2.2 & 1.37 & 61.67 & 173.64 & 1055.33 & 1204.87 & 3 \\
J143921.24+382149.0 & 219.83854 & 38.363624 & 0.17048 & 7.44 & 19.42339 & 18.60116 & 17.89776 & 17.53415 & 17.38088 & 2.46 & 1.39 & 1.06 & 62.92 & 115.87 & 693.48 & 482.81 & 3 \\
J023348.39+041816.0$^b$ & 38.4516497 & 4.3044494 & 0.16708 & 19.46 & 18.5291 & 18.2696 & 17.65791 & 17.14697 & 17.18564 & 2.85 & 0.78 & 0.99 & 39.57 & 480.73 & 260.74 & 133.51 & 4\\
J111338.60+552440.6 & 168.410867 & 55.4113 & 0.03778 & 8.86 & 17.78666 & 16.42867 & 15.71975 & 14.96744 & 14.99021 & 2.21 & 2.18 & 1.1 & 217.56 & 549.15 & 1545.07 & 902.65 & 4 \\
J150229.79+000741.6$^a$ & 225.624149 & 0.128245 & 0.08683 & 25.25 & 18.9067 & 17.15243 & 16.49912 & 16.21316 & 16.01477 & 3.27 & 2.36 & 1.33 & 9.49 & 70.41 & 213.53 & 269.61 & 5 \\
J081207.05+063355.6 & 123.02941 & 6.56545 & 0.11109 & 8.45 & 20.11894 & 18.87075 & 18.19683 & 17.80825 & 17.53576 & 2.32 & 1.9 & 1.15 & 22.88 & 12.08 & 193.64 & 87.48 & 5 \\
		\hline
	\end{tabular}
	\begin{tablenotes}
        %\tiny
        %\footnotesize
        \scriptsize
        \item[1]Notes: Columns (1)--(5): Designation, Ra, Dec, z(redshift) and S/N\_r(signal-to-noise ratio in r band) retrieved from the LAMOST catalogue; Columns (6)--(10): Petrosian magnitudes in u, g, r, i and z band retrieved from the photometric catalogue of SDSS; Column (11)--(13) represent correlation with morphological types, where Column (11) is the concentration index which is computed using petroR90, petroR50 in the $r$ band cross-matched photometric catalogue of SDSS, Column (12) is the (u - r) color which is computed using magnitudes of u and r  band also from SDSS, Column (13) is the 4000\AA \ break which is computed using the narrow definition D4000$_n$ introduced by \cite{Balogh et al1999}; Column (14)--(17) are line fluxes( H$\beta$, [OIII]$\lambda$5007, H$\alpha$ and [NII]$\lambda$6585) which are used in our work; Column (18) is the spectral class using our classification scheme, and the number 0, 1, 2, 3, 4, 5 represent passive, star-forming, composite, LINER, Seyfert, H$\alpha$-weak respectively. Records are marked with `$ a $' at the upper right corner of the designation if the spectra are from a subsample in the Northern Galactic Cap which includes 2,859 spectra complementary to SDSS main galaxy sample in the study of galaxy pairs. While records marked with `$ b $' represent the spectra targeted from the LAMOST Complete Spectroscopic Survey of Pointing Area at Southern Galactic Cap, including 4,493 spectra.
      \end{tablenotes}
\label{tab:catalog_Main}
\end{table*}

\section{Composite spectra}
\label{sec:composite_spec}

It is generally known that the composite spectra are of high S/N, which are suitable to detect spectral features. And we can also use composites as galaxy classification templates. In this section, we calculate the composite spectra of different classes to analyze the global continuum and spectral features, and further explore what spectral features are sensitive to different classes. We use median calculation to generate the composite spectra, because median spectrum can preserve the relative fluxes of the emission features and not alter the emission-line ratios (\citealt{Vanden Berk et al2001}), which is very important for classification.

\subsection{Constructing the Composite spectra}
\label{sec:composite_calcu}

We select spectra in our Main sample using the redshift and absolute magnitude cuts in order to eliminate the evolution effect and Malmquist bias to the composite spectra. The redshift and magnitude ranges for each class in our volume-limited sample are listed in Table \ref{tab:composite_criteria}. The redshift and absolute magnitude ranges are the same as \cite{Dobos et al2012}.

\begin{table}
\renewcommand\arraystretch{1.5}
	\centering	
	\caption{Redshift and absolute magnitude cuts for each galaxy class in our sample to calculate the composite spectra.}
\label{tab:composite_criteria}
\begin{tabular}{ccccc}
\hline
 & $z_{min}$ & $z_{max}$ & $M_{r,min}$ & $M_{r,max}$ \\
\hline
passive & 0.03 & 0.14 & -20.5 & -21.5 \\
H$\alpha$-weak & 0.03 & 0.14 & -20.5 & -21.5 \\
SF & 0.03 & 0.08 & -19.2 & -21.0 \\
Composite & 0.03 & 0.14 & -20.5 & -21.5 \\
LINER & 0.04 & 0.17 & -21.0 & -22.0 \\
Seyfert & 0.03 & 0.14 & -20.5 & -21.5 \\
\hline
\end{tabular}
\end{table}

\begin{figure*}
	\includegraphics[width=16.5cm]{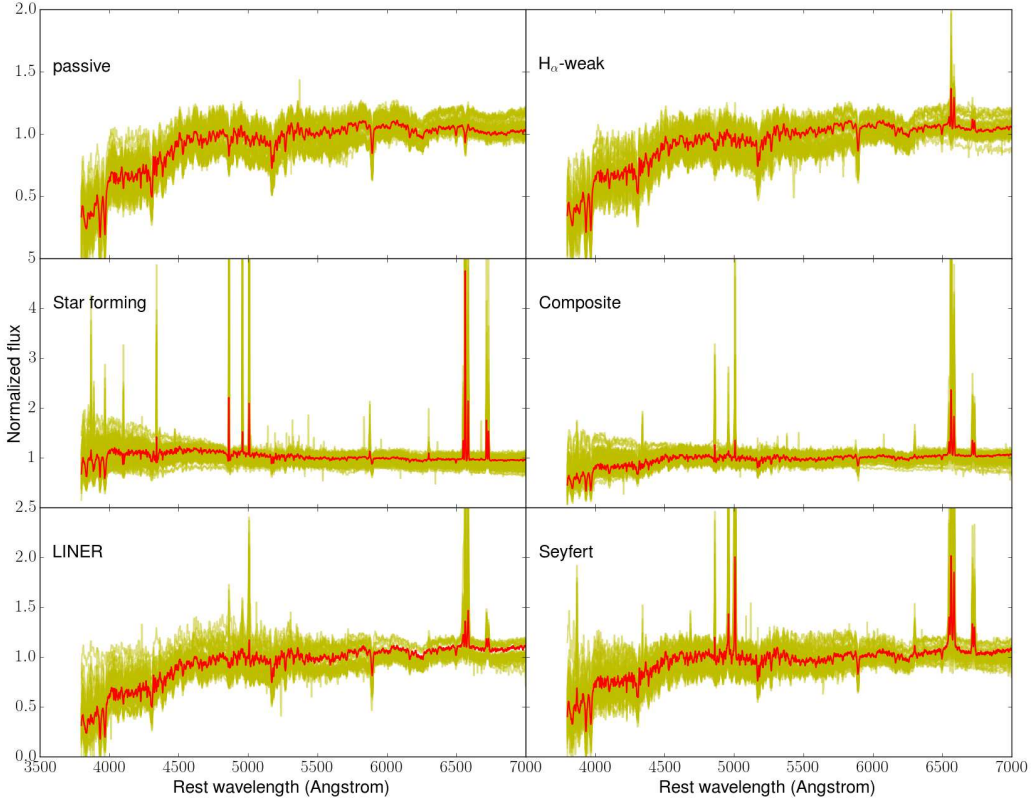}
    \caption{Composite spectra of different classes in our galaxy sample. In each panel, the red spectrum is the composite of galaxies in this class, and the yellow shaded spectra represent the 0.5$\sigma$ variation about the composites.}
    \label{fig:average_spec}
\end{figure*}

We pre-process the spectra before calculating the composites of different classes. First, the spectra are normalized by the median flux in the  4600-4800\AA \ region in which there are no strong emission lines. And then spectra are shifted to rest frame and rebinned to 1\AA. The final median composite spectra are plotted for different classes in Figure \ref{fig:average_spec}. In each panel, the yellow shaded spectra are random selected spectra from the class and the red spectrum is the composite spectrum of galaxies in this class. From this figure, the composite spectrum for passive has no emission lines totally, and the composite spectrum for H$\alpha$-weak has weak H$\alpha$ emission, no H$\beta$, [OIII], and [NII] emission lines. The other four composite spectra of emission-line classes have emissions with different intensities. In general, the composite spectrum for LINER is similar to the one of H$\alpha$-weak except for the [OIII] emission presence in LINER's composite spectrum. In fact some `retired galaxies' (RGs) lie in the region of LINER on BPT diagram \citep{Cid Fernandes et al2011}, i.e. galaxies that have stopped forming stars and are ionized by their hot low-mass evolved stars. \citealt{Cid Fernandes et al2011} deemed that RGs and passive galaxies are very similar objects.

The error of the median composite spectrum is calculated using the method of \citealt{Vanden Berk et al2001}, which is computed by dividing the 68\% semi-interquartile range of the flux densities by the square root of the number of spectra contribution to each bin. The wavelength, flux and uncertainty of flux of median spectra are available online.

\subsection{The continuum}
\label{sec:composite_conti}

We fit the continua of composite spectra with population synthesis models of BC03 via the same method described in Section \ref{sec:line_m_process}. For each composite spectrum, a best-fitting model is determined by best linear combination of models. Using the best-fitting models, we extend the wavelength coverage of our composite spectra in order to avoid the limited wavelength coverage because of the redshift range of our spectra sample. With these extended composites, we can explore more global features. And we can also use these composites with wider wavelength as templates for spectral classification in LAMOST.

\subsection{Emission and absorption lines}
\label{sec:composite_em}
The high S/N of the composites allows us to detect emission and absorption features. Most emission lines and absorption lines are identified in the composite spectra. We present the emission-line equivalent widths of all composites in Table \ref{tab:em_composite}, which are measured using the method described in Section \ref{sec:line_m_process} and the continua fits in Section \ref{sec:composite_conti}. The BPT diagram for the composites of the star-forming galaxies, composite galaxies, LINERs and Seyferts are in Figure \ref{fig:bpt_em} with red star markers. The four markers locate their corresponding classes, which suggests the median method for composites does not alter the line ratios.

\begin{table}
%\tiny
%\footnotesize
\scriptsize
\caption{Equivalent widths (\AA) of emission lines of six composite spectra.}
\label{tab:em_composite}
\begin{tabular}{ccccccc}
\hline
  & passive & H$\alpha$-weak & SF & Composite & LINER & Seyfert \\
\hline
OII3727 & -- & 3.618 & 17.025 & 8.787 & 5.704 & 11.036 \\
H$\delta$ & -- & -- & 2.712 & 1.469 & -- & -- \\
H$\gamma$ & -- & -- & 3.946 & 1.936 & -- & -- \\
H$\beta$ & -- & -- & 2.712 & 1.469 & 0.454 & 1.049 \\
OIII5008 & -- & -- & 3.396 & 2.169 & 1.795 & 6.917 \\
OI6302 & -- & -- & 1.181 & 0.825 & 0.794 & 1.146 \\
NII6548 & -- & 0.461 & 2.483 & 1.947 & 1.243 & 2.405 \\
H$\alpha$ & -- & 2.101 & 32.162 & 13.055 & 3.528 & 10.344 \\
NII6584 & -- & 1.326 & 7.456 & 5.621 & 2.99 & 6.54 \\
SII6717 & -- & 0.959 & 5.464 & 2.711 & 1.753 & 2.771 \\
SII6731 & -- & 0.9 & 3.935 & 2.183 & 1.569 & 2.422 \\
\hline
\end{tabular}
\end{table}

We calculate absorption-line indices of the composite spectra, which are summarized in Table \ref{tab:indices_composite}. These indices are adopted from Lick indices (\citealt{Worthey et al1994}; \citealt{Worthey et al1997}), BH indices( \citealt{Huchra et al1996}), DTT indices( \citealt{Diaz et al1989}), and D4000$_n$( \citealt{Balogh et al1999}). The values for atomic indices are expressed in angstroms of equivalent width, while those for molecular indices are expressed in magnitude.

\begin{table}
%\tiny
\scriptsize
\caption{Values of indices of six composite spectra.}
\label{tab:indices_composite}
\begin{tabular}{p{1.0cm}p{0.4cm}p{0.7cm}p{1.1cm}p{0.7cm}p{0.8cm}p{0.75cm}p{0.75cm}}
\hline
  &   & passive & H$\alpha$-weak & SF & Composite & LINER & Seyfert \\
\hline
Ca4227$^a$ & \AA & 1.253 & 1.165 & 0.533 & 0.712 & 1.12 & 0.843 \\
G4300$^a$ & \AA & 6.104 & 5.582 & 1.391 & 3.082 & 5.369 & 4.365 \\
Fe4383$^a$ & \AA & 5.102 & 4.721 & 1.814 & 3.052 & 4.991 & 4.271 \\
Ca4455$^a$ & \AA & 1.666 & 1.608 & 0.693 & 1.088 & 1.546 & 1.265 \\
Fe4531$^a$ & \AA & 3.598 & 3.447 & 2.026 & 2.693 & 3.437 & 3.093 \\
C4668$^a$ & \AA & 6.409 & 6.021 & 1.965 & 3.776 & 6.144 & 4.212 \\
H$\beta$ $^a$ & \AA & 1.944 & 1.403 & -4.322 & -1.223 & 0.728 & -0.685 \\
Fe5015$^a$ & \AA & 4.739 & 4.429 & -0.805 & 1.974 & 3.372 & -9.322 \\
Mg$_b$ $^a$ & \AA & 3.784 & 3.582 & 1.538 & 2.706 & 4.038 & 3.128 \\
Fe5270$^a$ & \AA & 2.953 & 2.909 & 1.835 & 2.321 & 2.806 & 2.599 \\
Fe5335$^a$ & \AA & 2.566 & 2.525 & 1.686 & 2.114 & 2.444 & 2.372 \\
Fe5406$^a$ & \AA & 1.584 & 1.534 & 0.946 & 1.239 & 1.536 & 1.414 \\
Fe5709$^a$ & \AA & 0.92 & 0.912 & 0.619 & 0.78 & 0.898 & 0.872 \\
Fe5782$^a$ & \AA & 0.758 & 0.766 & 0.55 & 0.717 & 0.789 & 0.752 \\
NaD$^a$ & \AA & 3.375 & 3.249 & 1.225 & 2.769 & 3.459 & 2.593 \\
H$\delta_A$ $^a$ & \AA & -1.572 & -0.706 & 3.422 & 1.998 & -0.874 & 0.981 \\
H$\gamma_A$ $^a$ & \AA & -4.097 & -3.368 & 0.199 & -0.861 & -4.065 & -2.27 \\
H$\delta_F$ $^a$ & \AA & 0.724 & 1.03 & 2.278 & 2.017 & 0.918 & 1.593 \\
H$\gamma_F$ $^a$ & \AA & -0.781 & -0.596 & 0.213 & 0.226 & -0.962 & -0.38 \\
CaII8498$^c$ & \AA & 1.3 & 1.342 & 1.355 & 1.529 & 1.736 & 1.538 \\
CaII8542$^c$ & \AA & 3.433 & 3.504 & 3.361 & 3.738 & 4.366 & 3.832 \\
CaII8662$^c$ & \AA & 2.883 & 3.034 & 3.087 & 3.651 & 4.585 & 3.212 \\
MgI8807$^c$ & \AA & 0.488 & 0.496 & 0.302 & 0.434 & 0.561 & 0.393 \\
CN$_1$ $^a$ & mag & 0.052 & 0.033 & -0.052 & -0.029 & 0.045 & -0.006 \\
CN$_2$ $^a$ & mag & 0.093 & 0.073 & -0.028 & 0.009 & 0.081 & 0.03 \\
Mg$_1$ $^a$ & mag & 0.099 & 0.093 & 0.026 & 0.056 & 0.101 & 0.067 \\
Mg$_2$ $^a$ & mag & 0.239 & 0.225 & 0.089 & 0.149 & 0.234 & 0.172 \\
TiO$_1$ $^a$ & mag & 0.028 & 0.026 & 0.012 & 0.018 & 0.027 & 0.021 \\
TiO$_2$ $^a$ & mag & 0.066 & 0.064 & 0.04 & 0.052 & 0.068 & 0.059 \\
CNB$^b$ & mag & 0.241 & 0.204 & 0.058 & 0.104 & 0.194 & 0.096 \\
H+K$^b$ & mag & 0.387 & 0.349 & 0.192 & 0.246 & 0.344 & 0.27 \\
CaI$^b$ & mag & 0.015 & 0.014 & 0.003 & 0.008 & 0.011 & 0.008 \\
G$^b$ & mag & 0.255 & 0.236 & 0.078 & 0.143 & 0.23 & 0.183 \\
H$\beta$$^b$ & mag & 0.069 & 0.06 & -0.037 & 0.018 & 0.045 & 0.023 \\
MgG$^b$ & mag & 0.101 & 0.096 & 0.034 & 0.068 & 0.109 & 0.079 \\
MH$^b$ & mag & 0.059 & 0.053 & 0.008 & 0.027 & 0.048 & -0.018 \\
FC$^b$ & mag & 0.07 & 0.066 & 0.037 & 0.048 & 0.068 & 0.054 \\
NaD$^b$ & mag & 0.081 & 0.076 & 0.017 & 0.056 & 0.079 & 0.053 \\
D4000$_n$ & -- & 1.863 & 1.229 & 1.392 & 1.752 & 1.476 & 1.732 \\
\hline
\end{tabular}
\begin{tablenotes}
\scriptsize
\item[1]Notes: indices$^a$ are Lick indices from \cite{Worthey et al1994} and \cite{Worthey et al1997}; indices$^b$ are from Brodie and Hanes (see \cite{Huchra et al1996}); indices$^c$ are from \cite{Diaz et al1989}; and D4000$_n$ is from \cite{Balogh et al1999}.
\end{tablenotes}
\end{table}

\begin{figure}
	\includegraphics[width=\columnwidth]{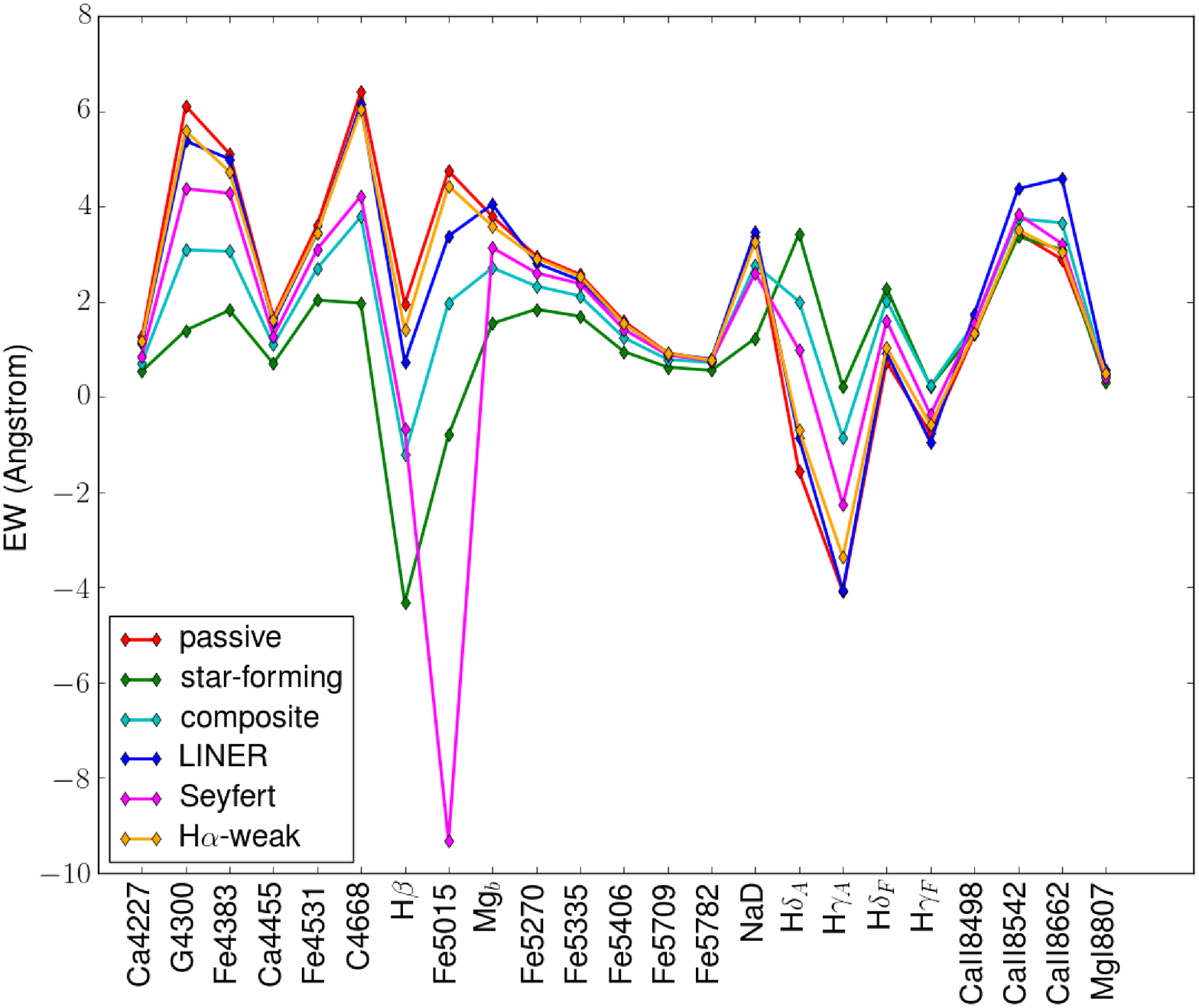}
	\includegraphics[width=\columnwidth]{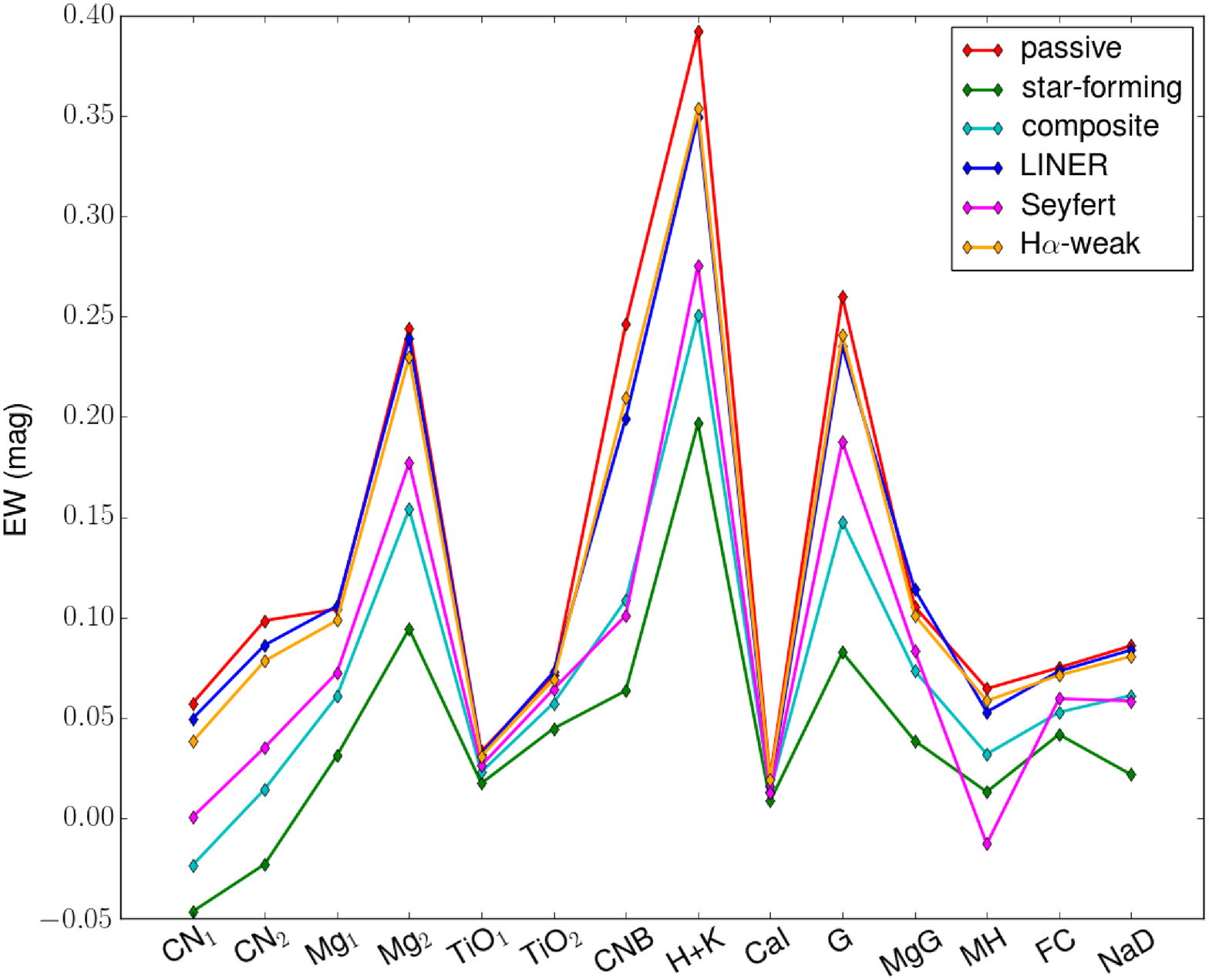}
    \caption{Indices of composite spectra which come from Table \ref{tab:indices_composite}. The indices in the upper panel are defined in \AA , while indices in the lower panel are defined in mag.}
    \label{fig:indics_composite}
\end{figure}

Using the absorption line indices, we can explore what other features can separate different classes of galaxies besides the four emission features we use in classification schema. Figure \ref{fig:indics_composite} shows the capability of line indices to differentiate the spectral class. It illustrates that some spectral lines are sensitive to the galaxy classes: G4300, Fe4383, H$\beta$,  Fe5015, Mgb, H$\delta_A$, H$\gamma_A$, which are measured as atomic absorption lines, and CN$_{1}$, CN$_{2}$, Mg$_{1}$, Mg$_{2}$, CNB, HK, G, MH, which are molecular bands. On one hand, these features are good class tracers so that they can be taken into account for spectral classification. On the other hand, these features can be used for comprehending the physical parameters of different classes such as age and metallicity. Fe5015, Mgb, Mg$_{1}$, Mg$_{2}$ are metal-sensitive indicators according to the definitions in \cite{Bruzual et al2003} and \cite{Thomas et al2003}, where three indices are defined as metal-sensitive indices: [Mg$_1$Fe], [Mg$_2$Fe] and [MgFe]$'$. The three indices of six composite spectra are calculated, which suggest that the metallicity of composite spectrum of star forming galaxies is the lowest, and that of passive galaxies is the highest. Age-sensitive indices H$\beta$, H$\delta_A$ and H$\gamma_A$ (\citealt{Bruzual et al2003},\citealt{Gallazzi et al2003})are significantly different for six galaxy classes in Figure \ref{fig:indics_composite}. The composite spectrum of star forming has the lowest value of H$\beta$ index, while the one of passive has the highest value, which suggests that the age of galaxies is older as H$\beta$ index is larger. Besides these indicators, there are some other lines(G band, CN$_{1}$, CN$_{2}$, CNB, HK, MH) with obvious difference among classes in Figure \ref{fig:indics_composite}, which are potential indicators for the study of stellar population.

\subsection{Comparison with composite spectra of SDSS}
\label{sec:comparison_sec}

\begin{figure}
	\includegraphics[width=\columnwidth]{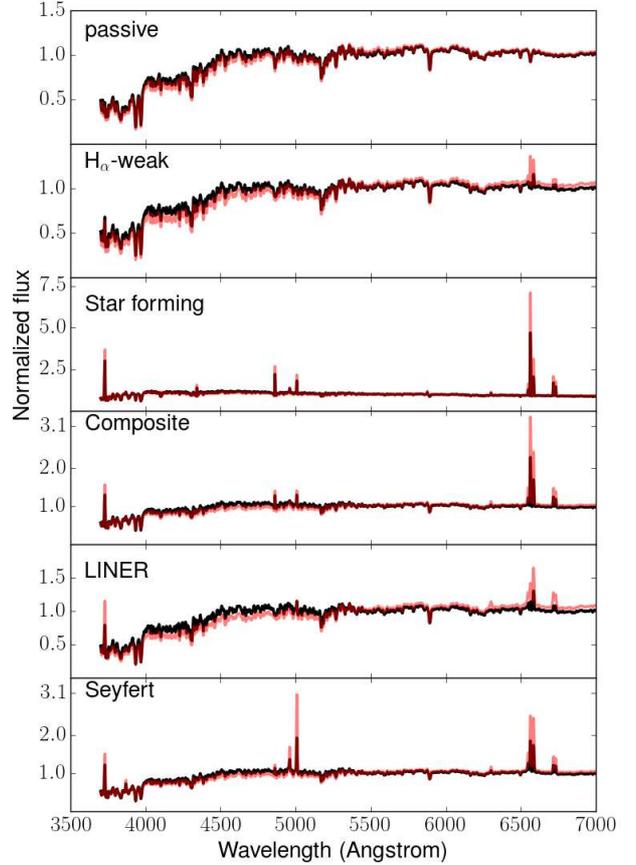}
    \caption{Comparison of LAMOST composite spectra with SDSS composites. The spectra in black are LAMOST and the spectra in red are SDSS.}
    \label{fig:average_comp}
\end{figure}

In \cite{Dobos et al2012}, they classified spectra of galaxies in SDSS DR7 into the six classes similar to ours, and also presented a set of composite spectra of different classes. In this section, we compare our composite spectra with theirs. The comparison of composites from LAMOST and SDSS is shown in Figure \ref{fig:average_comp}. The spectra in black are from LAMOST, and red are from SDSS. From this figure, the continua and spectral lines of the two sets of composites are slightly different.

\begin{figure}
	\includegraphics[width=\columnwidth]{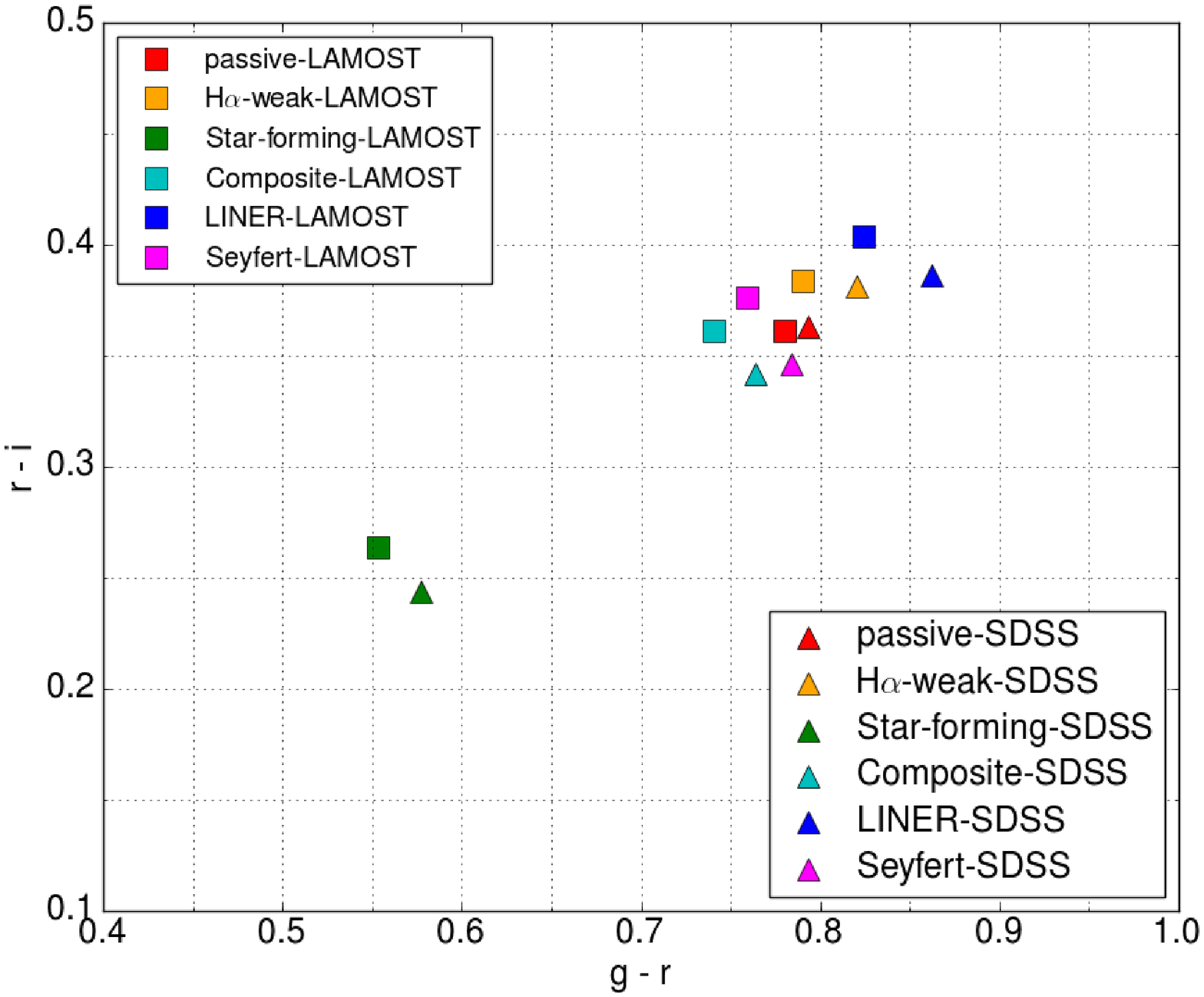}
    \caption{Color-color diagram of LAMOST composites and SDSS composites for the SDSS $ g-r $ and $ r-i $ color. The squares represent the loci of LAMOST composites and the triangles mark the loci of SDSS composites. The different colors of the squares and triangles represent different galaxy classes. \textbf{There exist 0.02 mag difference in $g-r$ and 0.01 mag difference in $r-i$ between LAMOST and SDSS composites because of calibration effects.}}
    \label{fig:compare_colors}
\end{figure}

The continua of LAMOST composites can be compared with the SDSS ones by color-color diagram. We compute the synthetic magnitudes of the composites in SDSS $ g $, $ r $, $ i $ bands. Figure \ref{fig:compare_colors} displays the comparison of the $ g-r $ and $ r-i $ color-color loci of LAMOST composites and SDSS composites. The composites of LAMOST spectra are bluer than those of SDSS by about 0.02 mag in $ g-r $ color, while in $ r-i $ color the LAMOST composites are redder than SDSS ones by about 0.01 mag. \textbf{These color differences mainly come from calibration effects.}

The relative difference of equivalent widths of main emission lines and absorption-line between LAMOST composites and SDSS ones is summarized in Table \ref{tab:compare_em_absor}. From the table, we can see the emission-line equivalent widths of LAMOST composites are less than those of SDSS within 40\%, and the absorption-line indices of LAMOST composites are less than those of SDSS within 10\%. These differences agree with the comparison between their averaging method for composites and median composites in \cite{Dobos et al2012} (seen Figure 15, 16 in their paper). In addition, on BPT diagram in Figure \ref{fig:bpt_em}, we plot SDSS composites of four classes: star-forming, composite, LINER and Seyfert with the cyan triangle markers. The differences of line ratios are small except for the composites of LINERs, which might be due to the different methods of classifying Seyferts and LINERs used to calculate the composite spectra.

\begin{table}
\newcommand{\tabincell}[2]{\begin{tabular}{@{}#1@{}}#2\end{tabular}}
%\tiny
\scriptsize
\caption{Relative difference of equivalent widths of emission lines and absorption-line indices between LAMOST composites and SDSS composites ((value\_LAMOST -value\_SDSS)/value\_SDSS)(in percent).}
\label{tab:compare_em_absor}
\begin{tabular}{p{0.85cm}p{0.78cm}p{0.63cm}p{0.83cm}p{0.7cm}p{0.85cm}*{2}{p{0.7cm}}}
\hline
 & & passive & H$\alpha$-weak & SF & Composite & LINER & Seyfert \\
\hline
\multirow{6}{*}{\tabincell{c}{Emission\\ lines}} & OII3727 & -- & -10.0 & -7.8 & -13.8 & -35.0 & -14.9 \\
& H$\beta$ & -- & -- & -18.7 & -26.4 & -1.8 & -36.7 \\
& OIII5008 & -- & -- & -20.5 & -30.9 & -2.0 & -37.7 \\
& NII6548 & -- & -- & -31.2 & -34.9 & -16.7 & -34.5 \\
& H$\alpha$ & -- & -33.4 & -29.9 & -25.2 & -36.4 & -38.4 \\
& NII6584 & -- & -26.5 & -31.1 & -30.2 & -38.4 & -38.5 \\
\hline
\multirow{7}{*}{\tabincell{c}{Absorption\\ lines}}& Fe5015 & -5.25 & -6.19 & -10.73 & 4.32 & 5.41 & -3.63 \\
& H$\delta_A$ & -4.9 & -11.19 & 0.45 & -10.41 & -11.69 & -10.82 \\
& H$\gamma_A$ & 3.91 & -5.34 & 8.21 & 7.76 & -10.29 & 5.88 \\
& Mg2 & -9.30 & -8.15 & -0.74 & -6.87 & -11.07 & -4.96 \\
& NaD & -10.3 & -9.43 & -8.88 & -10.33 & -11.67 & -7.96 \\
& HK & -3.26 & -3.94 & 2.72 & 8.76 & 0.31 & 7.72 \\
& D4000n & -1.36 & -2.32 & 2.96 & 5.33 & -0.05 & 6.70 \\
\hline
\end{tabular}
\end{table}

\section{Summary}
One goal of this paper is to provide a spectral classification of galaxies in LAMOST DR4 according to spectral line features and present a catalogue with these classification information and more accurate flux measurement of the nebular emission lines. From all spectra in LAMOST DR4, we focus on 40,182 spectra of 36,601 targets that have photometric information but no spectroscopic observations in SDSS DR13. Emission line is a key separation of galaxies, and accurate measurement of emission line intensities requires subtracting the best-fitting stellar population model. To overcome the error of population synthesis caused by the uncertainty of the continua, re-calibration of galaxy spectra in our sample is implemented through SDSS DR13 photometry. We then classify the galaxies into six classes: passive, H$\alpha$-weak, star-forming, composite, LINER and Seyfert based on four well measured lines: H$\beta$, [OIII]$\lambda$5007, H$\alpha$ and [NII]$\lambda$6585. A preliminary analysis for the results of classification is carried out, including statistical distributions, correlation with morphological types by three parameters: concentration index($ C $), ($ u $ - $ r $) color, D4000$_n$ index. A galaxy catalogue with classification information is provided. From the catalogue, we can also obtain the spectra of two special subsamples: a subsample in Northern Galactic Cap which includes 2,859 spectra complementary to SDSS main galaxy sample in the study of galaxy pairs, and the other subsample including 4,493 spectra targeted from The LAMOST Complete Spectroscopic Survey of Pointing Area at Southern Galactic Cap. In this work, we have presented a glimpse of classification of galaxies by spectral lines, and following work will bring more insight to physical properties of these different classes.

The other goal is to create a set of composite spectra for various galaxy classes from LAMOST spectra. The continua of the composite spectra are fitted with stellar population synthesis models to extend the wavelength coverage. From the spectral features of composites, we extract some features sensitive to classes such as H$\beta$, Fe5015, H$\gamma_A$, HK, and Mg$_{2}$ band, and investigate the correlation of some features with age and metallicity of each class. The comparison of our composite spectra with SDSS ones \citep{Dobos et al2012} indicates that they are roughly in agreement except for the emission line regions.

\section*{Acknowledgements}
We thank anonymous referees for valuable suggestions and comments. Thanks for helpful discussions from Du Bing toward the flux re-calibration and drawing Figure \ref{fig:compare_calib}. And thanks Budav\'{a}ri et al. for the helps about their code in \cite{Budavari et al.2009}. This work is supported by the National Key Basic Research Program of China (Grant No. 2014CB845700), and the National Natural Science Foundation of China (Grant Nos. 11390371 and 11233004).

The Guo Shou Jing Telescope (the Large Sky Area Multi-Object Fiber Spectroscopic Telescope, LAMOST) is a National Major Scientific Project built by the Chinese Academy of Sciences. Funding for the project has been provided by the National Development and Reform Commission. LAMOST is operated and managed by National Astronomical Observatories, Chinese Academy of Sciences.

%%%%%%%%%%%%%%%%%%%%%%%%%%%%%%%%%%%%%%%%%%%%%%%%%%



%\bibliographystyle{mnras}


%%%%%%%%%%%%%%%%%%%%%%%%%%%%%%%%%%%%%%%%%%%%%%%%%%

% Don't change these lines
\bsp	% typesetting comment
\label{lastpage}
\end{document}